\documentclass[%
 reprint,
 amsmath,amssymb,
 aps,
]{revtex4-2}

\usepackage{graphicx}
\usepackage{dcolumn}
\usepackage{bm}
\usepackage{caption}
\usepackage{upgreek}
\newcounter{mynumber}
\DeclareCaptionLabelFormat{nospace}{FIG~S#2}
\newcommand{\nextnumber}{\stepcounter{mynumber}\themynumber}

\begin{document}

\preprint{APS/123-QED}

\title{Polymer-modulated evaporation flow enables scalable \\self-assembly of highly aligned nanowires}

\author{Liyiming Tao$^{1,*}$}
\author{Zechao Jiang$^{2}$}
\thanks{These authors contributed equally to this work.}
\author{Shiyuan Hu$^{2}$}
\author{Lin Du$^{3,4}$}
\author{Qiuting Zhang$^{1}$}
\author{Jiajia Zhou$^{5,6}$}
\author{Masao Doi$^{7,8}$}
\author{Xiaojun Wu$^{3,4,9}$}
\author{Xingkun Man$^{2,10,\dagger}$}
\author{Ye Xu$^{1}$}

\email{manxk@buaa.edu.cn}
\email{ye.xu@buaa.edu.cn}

\affiliation{$^1$School of Mechanical Engineering and Automation, Beihang University, Beijing, 102206, China.}

\affiliation
{$^2$School of Physic, Beihang University, Beijing, 102206, China.}

\affiliation
{$^3$School of Electronic and Information Engineering, Beihang University, Beijing, 100191, China.}

\affiliation
{$^4$Hangzhou International Innovation Institute, Beihang University, Hangzhou, 311115, China.}

\affiliation
{$^5$South China Advanced Institute for Soft Matter Science and Technology, School of Emergent Soft Matter, South China University of Technology, Guangzhou, 510640, China.}

\affiliation
{$^6$Guangdong Provincial Key Laboratory of Functional and Intelligent Hybrid Materials and Devices, South China University of Technology, Guangzhou, 510640, China.}

\affiliation
{$^7$School of Physic, Beihang University, Beijing, 102206, China.}
\affiliation
{$^8$Wenzhou Institute, University of Chinese Academy of Science, Whenzhou, 325000, China.}

\affiliation
{$^9$Zhangjiang Laboratory, 100 Haike Road, Shanghai 201210, China.}

\affiliation
{$^{10}$Peng Huanwu Collaborative Center for Research and Education, Beihang University, Beijing, 100191, China.}

\date{\today}

\begin{abstract}
Highly aligned nanowire networks are essential for enabling anisotropic optical, electrical, and sensing functionalities in next-generation devices. 
However, achieving such alignment typically requires complex fabrication methods or high-energy processing. 
Here, we present a simple and scalable self-assembly strategy that uses a viscosity-enhancing polymer additive to modulate fluid flows during solvent evaporation. 
The addition of carboxymethylcellulose sodium (CMC-Na) reshapes the evaporation-driven flow field and generates a compressional flow region near the drying edge. 
Within this region, rotation-inducing velocity gradients progressively align silver nanowires (AgNWs) into highly ordered arrays.
This unique mechanism yields uniform AgNW coatings with a high degree of nanowire alignment and tunable areal density across centimeter-scale areas.
The resulting films exhibit strong broadband anisotropy, including polarization-dependent transmission in both visible and terahertz (THz) regimes and angle-dependent electrical conductivity. The approach also integrates naturally with dip-coating–based shear alignment, enabling programmable control over alignment direction and spatial patterning. This work establishes a robust, polymer-enabled mechanism for bottom-up nanowire alignment and offers a passive, energy-efficient route for fabricating anisotropic nanostructured coatings.
\begin{description}
\item[Keywords]
Nanowire alignment, Evaporation-induced flow, Anisotropic property
\end{description}
\end{abstract}

\maketitle


\section{Introduction}

One-dimensional (1D) nanomaterials such as nanowires and nanorods possess unique optical, electrical, and magnetic anisotropy due to their high aspect ratios, making them valuable for flexible electronics, solar cells, sensors, and photonic devices \cite{electronicskin1,electronicskin2,electronicskin3,electronicskin4,solarcell1,solarcell2,solarcell3,solarcell4,polarizer1,polarizer2}. 
However, these directional properties can only be exploited at device scale if the nanomaterials are uniformly aligned; otherwise, random orientation in disordered assemblies averages out their anisotropy \cite{missanisotropy1,missanisotropy2}. 
For emerging technologies that rely on directional transport or polarization-selective behavior, achieving tunable and scalable alignment of highly oriented 1D nanomaterials is essential \cite{polarization1,polarization2}.

A wide range of strategies have been developed to align 1D nanomaterials, including Langmuir–Blodgett assembly \cite{LB1,LB2,LB3}, co-assembly \cite{coassembly1,coassembly2,coassembly3}, solution shearing \cite{solutionshearcoating1,solutionshearcoating2,solutionshearcoating3,solutionshearcoating4}, dip-coating \cite{dip-coating1,dip-coating2,dip-coating3}, and microfluidic flow \cite{microfluidflow1,microfluidflow2,microfluidflow3}. 
These techniques typically rely on external fields, substrate motion, or surface confinement to induce alignment, but often involve complex fabrication setups, surface modifications, or mechanical actuation. 
Moreover, they tend to constrain the alignment direction, with dip-coating commonly producing nanowire orientation parallel to the withdrawal direction.

Evaporation-induced self-assembly offers a passive and scalable alternative. In such systems, flows driven by solvent evaporation—such as capillary or Marangoni flows—can guide nanowires toward the contact line, promoting alignment \cite{evaporationcapillaryflow, evaporationmarangoniflow}. 
Additional effects such as the “coffee-ring” phenomenon and crowding-induced entropy maximization near the drying edge can further enhance ordering, sometimes yielding liquid crystal–like structures \cite{evaporationassembly1, evaporationassembly2, evaporationassembly3, evaporationassembly4, evaporationassembly5, evaporationassembly6, evaporationassembly7}. 
However, despite these promising behaviors, evaporation-driven methods still lack reliable control over alignment direction, density, and uniformity. 
This limitation arises largely from insufficient understanding of the underlying flow mechanisms, and has hindered broader application of these methods in device-scale nanowire alignment.

Here, we introduce a new alignment mechanism based on polymer-modulated evaporation flow. 
By incorporating a viscosity-enhancing polymer, i.e. carboxymethylcellulose sodium (CMC-Na), into silver nanowire (AgNW) suspensions, we modulate the internal flow field during solvent evaporation. 
This adjustment induces a compressional flow region near the contact line, which systematically rotates nanowires and aligns them parallel to the drying front—a marked contrast to the perpendicular alignment typically produced by shear-driven dip-coating. 
The mechanism, elucidated through in situ microscopy and theoretical modeling, enables continuous deposition of highly aligned nanowires across a broad range of concentrations. 
The resulting AgNW coatings exhibit optical and electrical anisotropy and can be spatially patterned over centimeter-scale areas. 
It is also compatible with shear-based alignment: by controlling the contact line speed during dip-coating, the alignment direction can be dynamically tuned. 
These capabilities demonstrate a robust and versatile approach to bottom-up nanowire alignment, with potential applications in flexible electronics, photonic devices, and anisotropic sensing platforms.

\begin{figure*}[th!]
\centering
\includegraphics[width=0.95\textwidth]{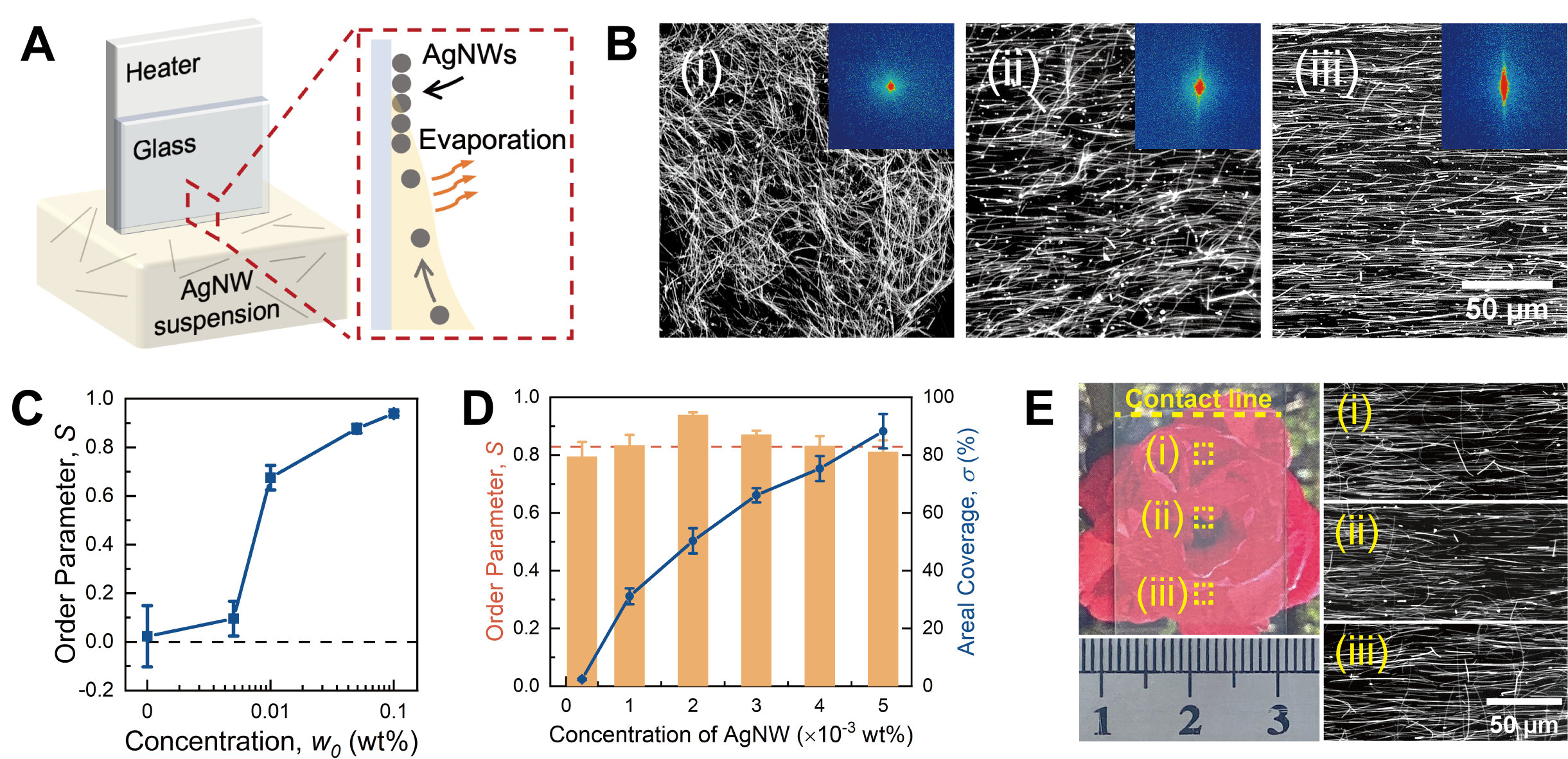}
\caption{\textbf{Fabrication of aligned AgNW coatings via evaporation-induced self-assembly.} 
(A) Schematics illustrating the alignment of AgNWs through evaporation-induced self-assembly within the dip-coating setup. 
(B) CRM images of AgNW coatings prepared from suspensions with different concentrations of CMC-Na. The concentrations of CMC-Na are set at 0 wt$\%$ (i), 0.01 wt$\%$ (ii), and 0.1 wt$\%$ (iii), respectively. Insets show corresponding light scattering patterns, with color indicating scattered laser intensity (red: highest; blue: lowest).
(C) The order parameter, $S$, calculated for AgNW coatings, plotted against the initial CMC-Na concentration $w_0$. 
(D) Order parameter, $S$, and areal coverage, $\sigma$, of AgNW coatings versus the initial concentration of AgNW $w_0$.
(E) Photograph of a transparent aligned AgNW coating prepared via dip-coating, along with CRM images from three distinct regions showing consistent alignment and density.}
\label{Fig. 1}
\end{figure*}

\section{Results and discussion}
\subsection{Controllable alignment of AgNWs in evaporating suspensions containing CMC-Na}

We first investigated AgNW alignment in a conventional dip-coating setup (FIG.~\ref{Fig. 1}A), where a glass slide, heated via a ceramic plate, was immersed in suspensions containing $2\times10^{-3}$ wt$\%$ AgNWs with CMC-Na concentrations ranging from 0 to 0.1 wt$\%$ (see Materials and Methods section for more details). 
As the solvent evaporated, capillary flow transported AgNWs to the contact line, depositing them onto the surface of the glass slide. 
Without CMC-Na, the deposited AgNWs formed a random network, as shown in the confocal reflection microscopy (CRM) image in FIG.~\ref{Fig. 1}B-i. 
With increasing CMC-Na concentration ($w_0 = 0.01$ wt$\%$ and $0.1$ wt$\%$), the nanowires aligned progressively parallel to the contact line (FIG.~\ref{Fig. 1}B-ii, iii), as further evidenced by the directional light scattering patterns in the figure insets. 
It is noteworthy that the alignment orientation of AgNWs achieved here contrasts with the \textit{perpendicular} alignment relative to the contact line typically observed in 1D nanomaterials deposited by traditional dip-coating methods, a phenomenon attributable to the shear flow effects near the substrate~\cite{dip-coating1,dip-coating2,dip-coating3}.

To quantify AgNW alignment, we analyzed nanowire orientation $\phi$ using OrientationJ, an ImageJ plugin, with $\phi = 0^\circ$ defined as the direction of the contact line (FIG.~S\nextnumber\label{Fig.S1 Distribution of angles}). 
The degree of alignment was measured using the nematic order parameter, $S = 2\langle \cos^2\phi \rangle - 1$, where $S = 0$ indicates random orientation and $S = 1$ denotes perfect alignment~\cite{orderedparameterdefine1,orderedparameterdefine2}. 
As shown in FIG.~\ref{Fig. 1}C, increasing the CMC-Na concentration to 0.1~wt\% raised the $S$ value from near zero to approximately 0.9, indicating a high degree of alignment. 
These results demonstrate that adding CMC-Na enables effective control over nanowire orientation.

We further examined how the concentration of AgNWs in the suspension influences the resulting coating. 
As shown in FIG.~\ref{Fig. 1}D and FIG.~S\nextnumber, increasing the AgNW concentration from $0.25 \times 10^{-3}$ to $5 \times 10^{-3}$~wt\% leads to a corresponding increase in areal coverage, $\sigma$, from 2\% to approximately 90\%. 
Meanwhile, the degree of alignment, quantified by the order parameter $S$, remains consistently high, \textit{i.e.}, $S > 0.8$, indicating that AgNW coatings with tunable density can be produced without compromising alignment quality. 

In addition to precise alignment and density control, our method enables large-scale fabrication of uniform AgNW coatings and films.  
A photograph in FIG~\ref{Fig. 1}E shows a well-aligned AgNW film with high transparency and uniformity over a centimeter-scale area.  
Microscopic CRM images from three distinct regions confirm consistent nanowire density and alignment across the entire $2 \times 2.5$~cm$^{2}$ area.  
We also demonstrate the fabrication of free-standing aligned nanowire films using alternative viscosity-modifying polymer systems.  
For example, N-methylpyrrolidone (NMP) solution containing Polyamic acid (PAA) and AgNWs can be used to produce flexible composite films with highly aligned AgNWs (see Appendix and FIG.~S\nextnumber).  
This versatility, combined with scalability and uniform alignment, positions our approach as a promising route for integrating 1D nanostructures into electronic, photonic, and flexible device platforms.

\subsection{Mechanism of AgNW alignment in evaporating suspensions containing CMC-Na}\label{subsec2.2}

\begin{figure*}[t!]
\centering
\includegraphics[width=0.95\textwidth]{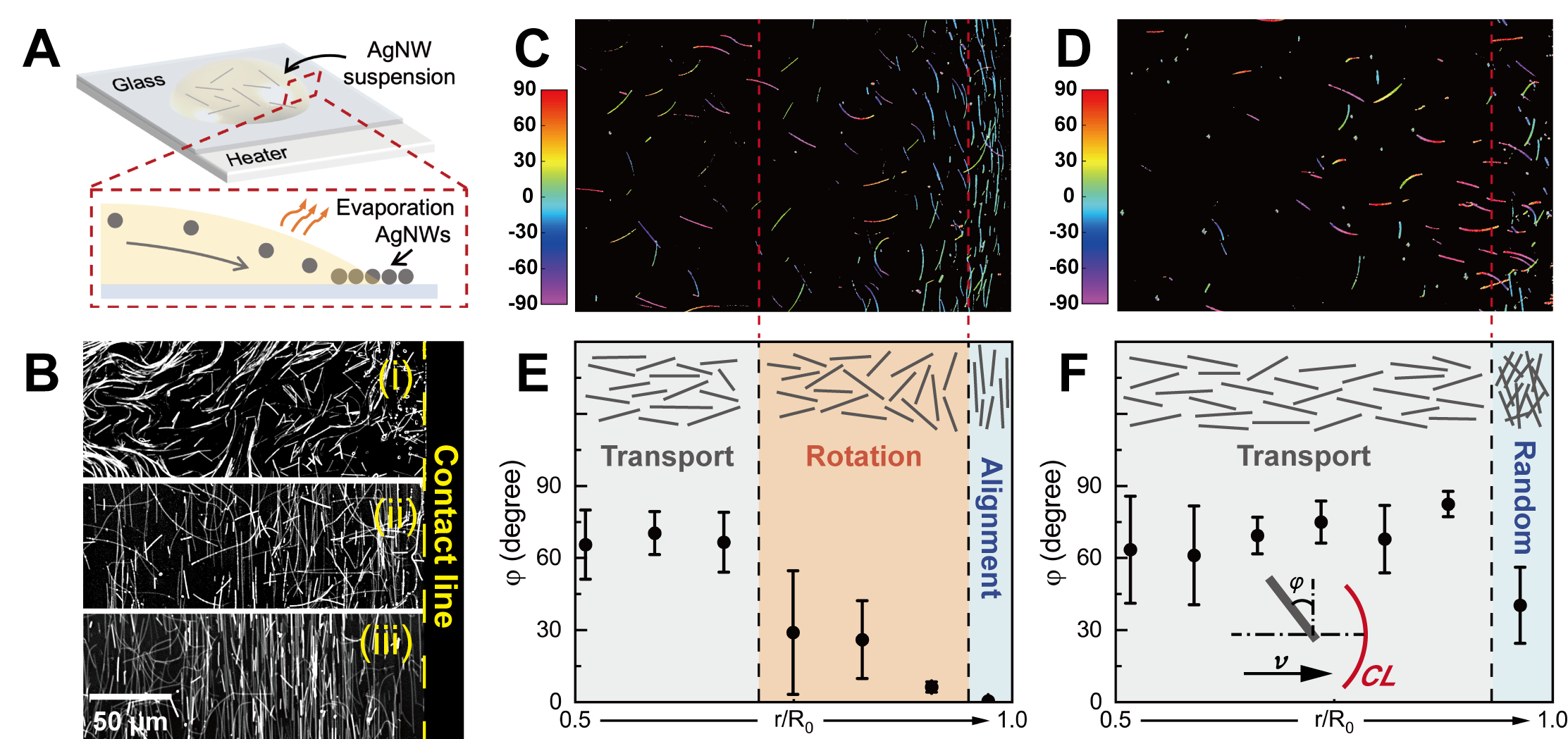}
\caption{
\textbf{Transport and deposition of AgNWs by the evaporation-induced flow in drying droplets.} 
(A) Schematics illustrating the alignment of AgNWs through evaporation-induced self-assembly within the droplet evaporation setup.
(B) CRM images of AgNW deposits near the contact line of evaporating droplets with different concentrations of CMC-Na. The concentrations of CMC-Na are 0 wt$\%$ (i), 0.01 wt$\%$ (ii), and 0.1 wt$\%$ (iii), respectively. The yellow dashed line indicates the location and orientation of the contact line.
(C) and (D) Representative micrographs showing AgNW orientations near the contact line in droplets with (C) and without (D) CMC-Na. 
Nanowires are color-coded by the angle $\varphi$ between their long axis and the tangential direction of the contact line. 
(E) and (F) Angle $\varphi$ plotted as a function of normalized radius $r/R_0$ for droplets with (E) and without (F) CMC-Na. 
Inset: schematic showing the contact line, the direction of capillary flow ($V$), and the definition of angle $\varphi$.}
\label{Fig. 2}
\end{figure*}

To investigate the alignment mechanism, we tracked the dynamics of AgNW transport and deposition in drying droplets on glass substrates. 
Similar alignment behavior was observed in droplets containing CMC-Na, as shown in FIGs.~\ref{Fig. 2}A, \ref{Fig. 2}B, and FIG.~S\nextnumber.
In both dip-coating and drying-droplet setups, evaporation-induced flow governs AgNW orientation, with the flow field distribution playing a central role. 
Using time-lapse confocal microscopy, we captured nanowire motion near the contact line in two representative cases: one with 0.1~wt\% CMC-Na and one without. 
In the suspension containing CMC-Na, AgNWs were gradually carried outward and aligned parallel to the contact line upon deposition. 
In contrast, without CMC-Na, nanowires moved more rapidly toward the drying edge and were randomly deposited due to flow reflux near the edge.

To quantify the dynamics of AgNWs during deposition, we analyzed their positions and orientations frame-by-frame from the recorded time-lapse videos using OrientationJ. 
In two representative frames (FIGs.~\ref{Fig. 2}B and \ref{Fig. 2}C), individual AgNWs are color-coded by the angle $\varphi$ between their long axis and the tangential direction of the contact line. 
To visualize orientation changes during transport, we plotted $\varphi$ versus normalized radial position $r/R_0$, where $R_0 \approx 1200\,\mu$m is the initial droplet radius. 
Results averaged over eight frames taken from 30 to 100 seconds after the start of the drying process are shown in FIGs.~\ref{Fig. 2}E and \ref{Fig. 2}F. 
In suspensions with 0.1~wt\% CMC-Na, three spatial regions emerge: far from the edge, NWs align with $\varphi$ nearly 90$^\circ$ due to strong outward capillary flow; near the edge, $\varphi$ transitions gradually from 90$^\circ$ to 0$^\circ$, indicating orderly rotation; at the edge, most NWs align at approximately 0$^\circ$, consistent with the high final alignment observed in FIG.~\ref{Fig. 2}D-iii. 
In contrast, without CMC-Na, $\varphi$ changes abruptly from 90$^\circ$ to a wide range of values between 0$^\circ$ and 90$^\circ$, reflecting disordered deposition. 
This smooth rotational alignment when approaching the contact line appears unique to suspensions containing CMC-Na.

\begin{figure*}[t!]
\centering
\includegraphics[width=0.95\textwidth]{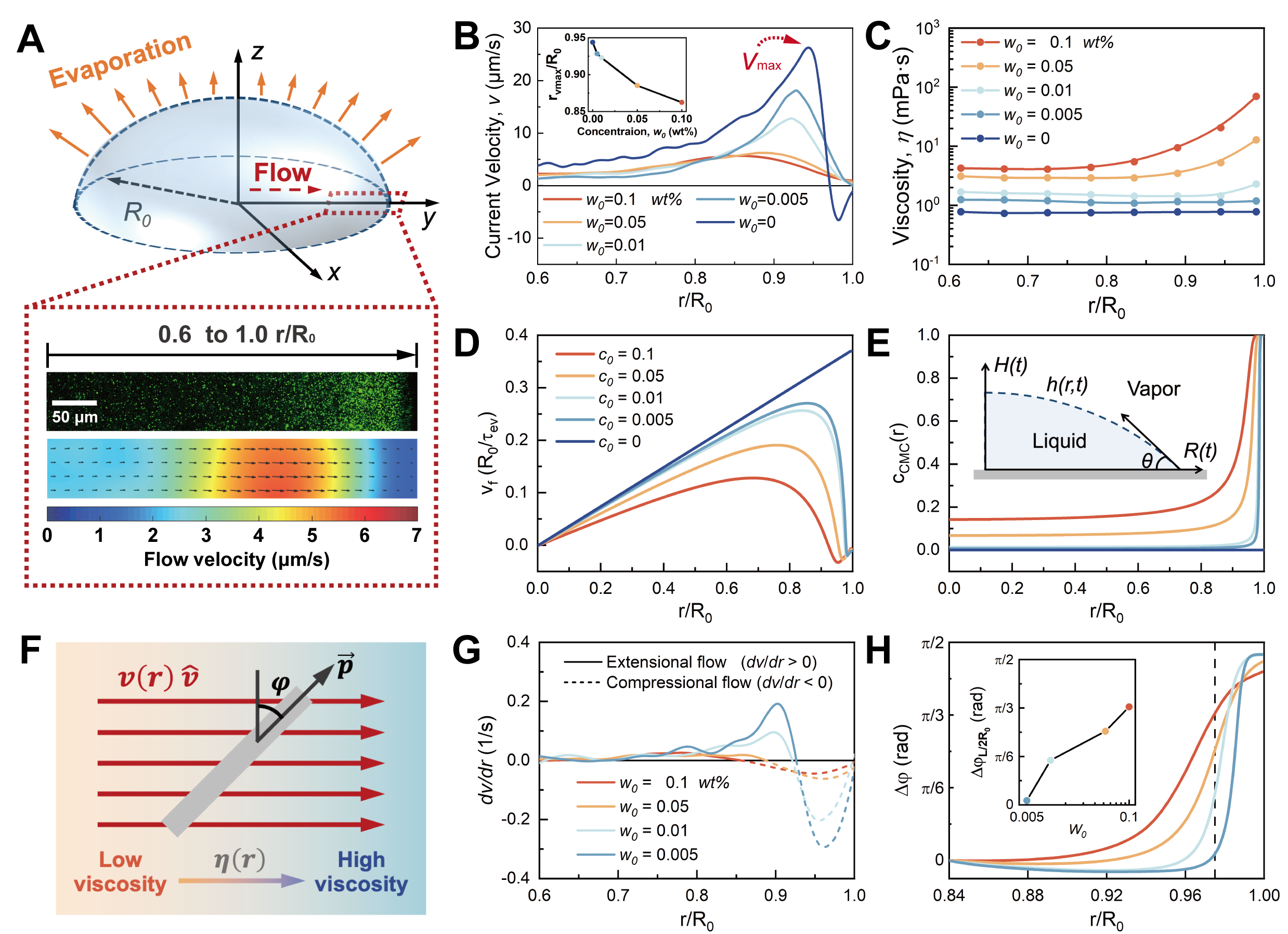}
\caption{\textbf{Experimental and theoretical analysis of the alignment mechanism of AgNWs in evaporating droplets containing CMC-Na.} 
(A) Schematic of the evaporating droplet setup for $\upmu$PIV analysis and a representative fluorescence image with the corresponding velocity distribution map for $w_0 = 0.1$ wt\%. 
(B) Radial flow velocity profiles, $v$, versus normalized radius, $r/R_0$, for droplets with varying CMC-Na concentrations during evaporation. \textit{Inset}: normalized position of peak velocity, $r_\text{vmax}/R_0$, as a function of initial CMC-Na concentration, $w_0$. 
(C) Local viscosity, $\eta$, as a function of $r/R_0$ for droplets with different initial CMC-Na concentrations.
(D) Theoretically calculated velocity profiles, $v_f$, versus $r/R_0$ at the characteristic evaporation time $t = 0.4 t_\text{total}$. 
(E) Radial distribution of CMC-Na concentration, $c(r)$, from theoretical modeling. \textit{Inset}: schematic of the droplet model showing height $H(t)$, contact angle $\theta$, and contact radius $R(t)$.
(F) Schematic illustration of nanowire rotation in flow due to velocity gradients.
(G) Velocity gradient, $dv/dr$, as a function of $r/R_0$, computed from the flow profiles. Solid and dashed lines indicate positive (extensional) and negative (compressional) gradients, respectively. 
(H) Angular change of nanowires, $\Delta\varphi$, versus $r/R_0$, based on experimental flow gradients. \textit{Inset}: $\Delta\varphi$ evaluated at a position half the nanowire length from the contact line, $r = R_0 - L/2$.
}
\label{Fig. 3}
\end{figure*}

Since AgNW motion is strongly influenced by fluid flow, we employed microparticle image velocimetry ($\upmu$PIV) to examine velocity distributions in drying droplets with varying CMC-Na concentrations, using fluorescent tracer particles (see Materials and Methods). 
Fluorescence imaging was conducted at a plane near the substrate across a normalized radial range of $r/R_0$ from 0.6 to 1.0, as illustrated in FIG.~\ref{Fig. 3}A. 
A representative fluorescence frame and the corresponding velocity map for $w_0 = 0.1$~wt\% are highlighted in the red dashed box. 
Additional $\upmu$PIV results for other CMC-Na concentrations are provided in FIG.~S\nextnumber\label{PIV Results}. 
Due to circular symmetry, flow was primarily radial, so we focused on the radial velocity $v$ as a function of $r/R_0$, shown in FIG.~\ref{Fig. 3}B. 
Typically, the evaporation-induced capillary flow increases radially outward, peaking before sharply dropping near the contact line. 
With CMC-Na, both the peak velocity $v_\text{max}$ and its location $r_\text{vmax}$ shift inward and decrease in magnitude, as seen in the inset of FIG.~\ref{Fig. 3}B. 
This flow field reveals two distinct regimes: an extensional region near the droplet center with increasing velocity, and a compressional region near the edge where velocity decays. 
The comparison among suspensions with different $w_0$ clearly demonstrated that increasing CMC-Na concentration expands the compressional region, indicating altered flow dynamics.

To better understand the CMC-Na-induced changes in flow behavior, we combined experimental measurements with theoretical modeling. 
We first measured that the evaporation rate of aqueous CMC-Na solutions decreases with increasing concentration from 0 to 1.0~wt\% under identical conditions (FIG.~S\nextnumber\label{ Evaporation Rate}), suggesting that CMC-Na slows local evaporation. 
We hypothesize that during drying, CMC-Na concentrates non-uniformly, especially near the droplet edge, affecting the local flow field. 
To probe this, we estimated the spatial distribution of CMC-Na using a semi-quantitative method that combines microrheology of fluorescent tracer particles with bulk viscosity measurements (see SI). 
The results indicate that in droplets containing CMC-Na, local viscosity near the droplet edge is consistently elevated relative to the interior.
In particular, for the 1.0~wt\% sample, the edge viscosity exceeds that of the droplet center by more than an order of magnitude (FIG.~\ref{Fig. 3}C). 
This pronounced viscosity gradient suggests substantial local enrichment of CMC-Na near the contact line (see Appendix and FIG.~S\nextnumber\label{Loacl Concentration of CMC-Na}). 
Such accumulation likely suppresses edge evaporation, contributing to the observed shift in flow pattern.

To rationalize the experimentally observed modification of flow patterns with CMC-Na, we developed a theoretical model for the height-averaged radial velocity \( v_f \) inside an evaporating droplet. 
As shown in FIG.~\ref{Fig. 3}E inset, we consider a droplet placed on a substrate and apply a model based on our previous work~\cite{theoricalmodelXKandDoi,jiang2025uniform}. 
The expression for \( v_f \) is derived from the mass conservation equation, incorporating a CMC-Na–dependent evaporation rate, a pinned contact line (\( \dot{R} = 0 \)), and a spatially varying solute concentration \( c(r,t) \):
\begin{equation}
\begin{aligned}
    v_f &= \frac{\pi D_{\text{ev}}^W rR}{4V}  
    - \frac{2\pi D_{\text{ev}}^W}{R} \left(\frac{r}{2V} 
    + \frac{r^3}{2\pi R^4 h}\right) \int_{0}^{R(t)} r c(r,t) dr \\
    &\quad + \frac{D_{\text{ev}}^W}{rRh} \int_{0}^{r} r' c(r',t) dr'.
\end{aligned}
\label{eq3}
\end{equation}

Here, \( h(r,t) \) is the local droplet height, \( R(t) \) and \( V(t) \) are the contact radius and droplet volume, \( D_{\text{ev}}^{W} \) is the diffusion coefficient of water in air, and \( c(r,t) \) is the height-averaged volume concentration of CMC-Na.

Equation~\ref{eq3} reveals that \( v_f \) is strongly influenced by the radial profile of \( c(r,t) \), which becomes non-uniform during evaporation due to the accumulation of polymeric solute near the contact line (FIG.~\ref{Fig. 3}E and Fig.~S\themynumber\label{Loacl Concentration of CMC-Na}). 
When the initial CMC-Na concentration in the droplet, denoted \( c_0 \), is zero, \( v_f \) increases linearly with radius. 
However, at non-zero \( c_0 \), \( v_f \) first rises, reaches a maximum, and then declines near the edge. 
As \( c_0 \) increases from 0.005 to 0.1, the location of this velocity peak shifts inward from \( 0.85R_0 \) to \( 0.7R_0 \), indicating an expansion of the compressional flow region. 
These theoretical predictions (FIG.~\ref{Fig. 3}D) align well with the experimental results in FIF.~\ref{Fig. 3}B, supporting the mechanistic interpretation provided by the model.
Further derivation details are available in the Supplementary Information.

To model nanowire rotation in flow regions with both velocity and viscosity gradients (FIG. \ref{Fig. 3}F), we apply resistive-force theory for slender bodies in fluids with a spatially varying viscosity~\cite{mason1995optical}.
We assume that the nanowires are short compared with the characteristic spatial scales of variations in both flow velocity and viscosity. 
Because the nanowires are passively advected by the flow, the total hydrodynamic force and torque acting on the nanowires are zero~\cite{NWsinfluid3}.
Enforcing these constraints, we obtain the angular velocity (see SI for detailed derivation):

\begin{equation}
\omega = \frac{d\varphi}{dt} = \frac{d v(r)}{dr}\sin\varphi \cos\varphi.
\label{Omega}
\end{equation}

Notably, $\omega$ depends solely on the velocity gradient and not on the viscosity profile.
Plotting $dv/dr$ vs. $r/R_0$ from data in FIG.~\ref{Fig. 3}B also reveals two distinct regions (FIG.~\ref{Fig. 3}G): extensional flow ($dv/dr > 0$), where nanowires rotate toward the flow direction ($\omega > 0$), and compressional flow ($dv/dr < 0$), where they rotate toward a perpendicular orientation ($\omega < 0$).

According to Equation~\ref{Omega}, the rotational speed also depends on the initial nanowire orientation $\varphi$ (see the inset in FIG.~\ref{Fig. 2}F). 
We numerically integrated this equation along radial trajectories with a step size of $\Delta r/R_0 = 0.001$ to compute the angular evolution $\Delta\varphi$ for droplets with different initial CMC-Na concentrations $w_0$. 
The resulting relationship between $\Delta\varphi$ and $r/R_0$ is plotted in FIG.~\ref{Fig. 3}H. 
Since AgNWs cease rotating upon deposition near the contact line, we evaluate the angular change $\Delta\varphi$ at a position located half the nanowire length from the edge, \textit{i.e.}, $r/R_0 = 1 - L/2R_0$ (FIG.~\ref{Fig. 3}H inset).
For $w_0 = 0.005\%$, the angular change $\Delta\varphi$ is nearly zero, while at $w_0 = 0.1\%$, $\Delta\varphi$ reaches up to $\pi/3$.
This enhanced nanowire rotation with increasing CMC-Na concentration reflects the broader and stronger compressional flow and explains the higher degree of alignment observed in AgNW arrays at elevated CMC-Na concentrations.

We would like to emphasize that the alignment mechanism in our system is fundamentally distinct from those based on entropy-driven assembly of closely packed nanorods near the contact line~\cite{evaporationassembly1, evaporationassembly2, evaporationassembly3, evaporationassembly4, evaporationassembly5, evaporationassembly6, evaporationassembly7}. 
Although such systems can also produce alignment parallel to the contact line, they rely on thermodynamic packing effects that require high particle density. 
In contrast, our method achieves alignment through polymer-mediated modulation of evaporation-induced flows, where the emergence of compressional flow fields actively rotates nanowires into ordered configurations.

A key advantage of this mechanism is its ability to decouple alignment quality from nanowire concentration. As previously shown in FIG.~\ref{Fig. 1}D, we achieve tunable areal coverage while maintaining high alignment ($S > 0.8$), a level of control not attainable with entropy-based methods. This flexibility enables the fabrication of AgNW coatings with independently optimized structural and optical properties.
Moreover, the method generalizes to diverse 1D nanomaterials, including silicon carbide and silicon dioxide nanowires (FIG.~S\nextnumber\label{Different NWs}), highlighting its broad applicability.

\begin{figure*}[t!]
\centering
\includegraphics[width=0.95\textwidth]{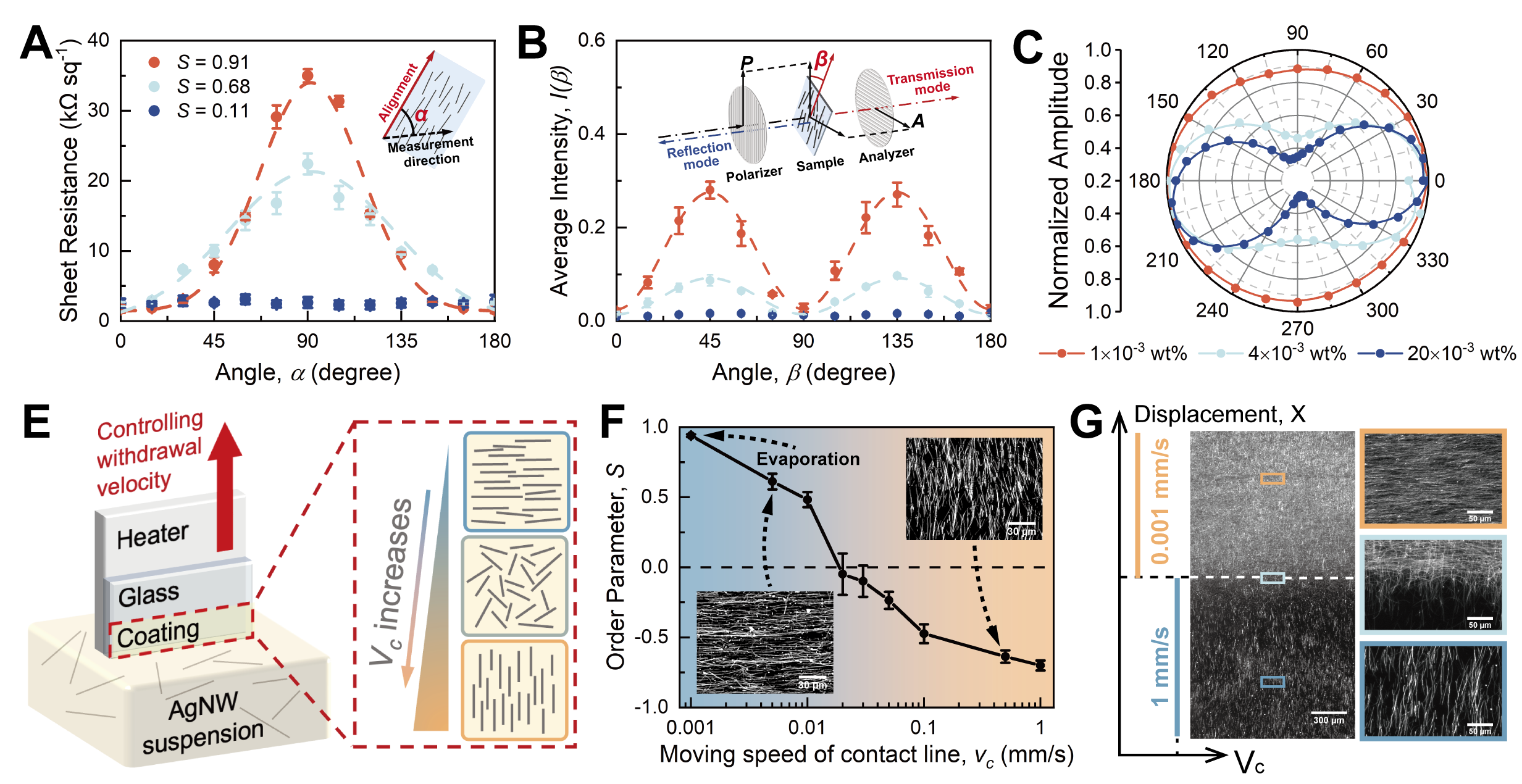}
\caption{\textbf{Characterization of anisotropic properties and programmable alignment in AgNW coatings.} 
(A) Sheet resistance as a function of measurement angle $\alpha$ relative to the alignment direction of the AgNW coating. 
Inset: schematic showing the definition of $\alpha$ in the four-point probe measurement setup.
(B) Normalized transmitted light intensity as a function of angle $\beta$ between the nanowire alignment and the polarizer axis. 
Dashed curves represent fits to Malus’s law. 
Inset: schematic of the optical measurement configuration with crossed polarizers and rotating sample stage.
(C) Normalized terahertz transmission amplitude as a function of the polarization angle relative to the alignment direction of AgNWs, demonstrating strong anisotropy in the THz regime.
(D) Schematic illustration of the combined alignment mechanisms: evaporation-induced compressional flow and shear flow from substrate withdrawal.
(E) Order parameter $S$ as a function of the contact line velocity $v_c$, showing a transition from parallel to perpendicular alignment. 
Insets: CRM images of AgNW coatings prepared at $v_c = 0.005$ mm/s and $v_c = 1$ mm/s.
(F) CRM images of AgNW coatings prepared using different programmed withdrawal modes, demonstrating alternating stripe patterns with orthogonal nanowire alignment.
}
\label{Fig. 4}
\end{figure*}

\subsection{Multispectral anisotropic properties of highly aligned AgNW coatings}

To evaluate the electrical anisotropy of the AgNW coatings, we measured the sheet resistance of three selected samples at various angles relative to the alignment direction, $\alpha$, as shown in FIG. \ref{Fig. 4}A. 
In well-aligned samples ($S = 0.91$), the sheet resistance varied strongly with measurement angle $\alpha$. 
The minimum resistance (1.60–1.72 k$\Omega$~sq$^{-1}$) was observed when the probe direction was parallel to the nanowire alignment ($\alpha = 0^\circ$ or $180^\circ$), while the maximum resistance ($\sim$34.6 k$\Omega$~sq$^{-1}$) occurred at $\alpha = 90^\circ$. 
In contrast, randomly oriented coatings showed no angle dependence, with resistance around 2.5 k$\Omega$~sq$^{-1}$ at all angles.
This pronounced anisotropy arises from both the directional arrangement of AgNWs and the presence of the insulating CMC-Na matrix, which limits conduction pathways perpendicular to the alignment. 
As demonstrated in FIG.~S\nextnumber, an LED connected across a well-aligned coating ($S = 0.90$) illuminates only when the voltage is applied parallel, not perpendicular, to the alignment direction.

The optical anisotropy of aligned AgNW coatings was characterized using polarized light microscopy. 
Samples were placed between crossed polarizers and rotated to vary the angle $\beta$ between the alignment direction and the polarization axis (FIG.~\ref{Fig. 4}B, inset). 
For well-aligned coatings ($S = 0.91$ and $S = 0.68$), the transmitted light intensity varied with $\beta$ and followed Malus's law~\cite{chartier2005introduction,hecht2012optics}, showing maximum brightness near $\beta = 45^\circ$ and a minimum at $0^\circ$ (FIG.~\ref{Fig. 4}B). 
In contrast, randomly oriented samples ($S = 0.11$) exhibited nearly constant intensity regardless of angle, confirming the absence of optical anisotropy.
Well-aligned AgNW coatings also exhibited distinct color variations under polarized reflection mode, as shown in FIG.~S\nextnumber.

Beyond the visible spectrum, highly aligned AgNW coatings exhibit strong polarization-dependent transmission in the terahertz (THz) regime.  
As shown in FIG.~\ref{Fig. 4}C, for samples with high areal coverage (95\%) and strong alignment ($S = 0.89$), the transmitted THz intensity varies markedly with polarization angle, dropping to just 20\% of the maximum when the electric field is oriented perpendicular to the nanowire alignment.  
This pronounced anisotropy reflects the directional conductivity of the aligned network.  
Samples with lower AgNW densities show reduced anisotropy, indicating that the THz polarization contrast also depends on the areal coverage of the coating.  
These results underscore the potential of densely packed, highly aligned AgNW coatings as polarizing components for terahertz photonic and communication devices.

\subsection{Programmable alignment of AgNWs through combined flow mechanisms}

We have shown that AgNWs can be aligned parallel to the contact line via evaporation-induced compressional flow. 
In a dip-coating setup, continuous evaporation allows the contact line to recede gradually, enabling deposition of large-area AgNW coatings aligned parallel to the contact line. 
In contrast, conventional dip-coating at high withdrawal speeds produces perpendicular alignment due to shear flow at the moving contact line.

We demonstrate that the alignment direction can be switched by varying the substrate withdrawal speed (FIG.~\ref{Fig. 4}D). 
The corresponding order parameter $S$ is plotted as a function of withdrawal speed in FIG.~\ref{Fig. 4}E. 
A clear transition in alignment direction is observed as the contact line speed increases. 
At very low speeds ($v_c = 0.001$ mm/s), alignment is parallel to the contact line ($S > 0$) due to compressional flow. 
Increasing the withdrawal speed introduces shear flow, which competes with the compressional flow. 
At $v_c = 1$ mm/s, shear flow dominates, leading to perpendicular alignment ($S < 0$).

This enables patterned coatings with alternating alignment directions, controllable via the withdrawal protocol. 
In FIG.~\ref{Fig. 4}F, a sample is prepared by evaporating a $5 \times 10^{-3}$ wt\% AgNW suspension with 0.1 wt\% CMC-Na at $75^\circ$C for 15 minutes. 
This is followed by a 1 mm withdrawal at $v_c = 1$ mm/s. 
The evaporation step yields a stripe of AgNWs aligned parallel to the contact line, while the withdrawal step induces perpendicular alignment, creating alternating stripes with orthogonal orientations. 
The width of each stripe can be tuned by controlling the timing of each step.
Another example (FIG.~S\nextnumber) shows alternating 15-minute evaporation periods and 1 mm withdrawals at $v_c = 10$ mm/s using a $2 \times 10^{-3}$ wt\% AgNW suspension. 
Fast withdrawal yields minimal deposition, forming 1 mm gaps. 
Repeating this two-step process enables fabrication of periodic AgNW patterns with controlled alignment.
These patterning capabilities offer significant potential for the fabrication of integrated micropatterned polarization sensors, smart windows, and other optoelectronic devices requiring spatially controlled anisotropy.

\subsection{Conclusion}\label{sec3}
In summary, we report a distinct alignment mechanism for one-dimensional nanomaterials, enabled by polymer-assisted modulation of evaporation-induced flow. 
By adding a viscosity-enhancing polymer like CMC-Na in AgNW suspensions, we demonstrate that local viscosity gradients give rise to compressional flow near the contact line, which in turn drives in-plane rotation and alignment of nanowires. 
This mechanism, confirmed through both experiments and theoretical modeling, is fundamentally different from traditional shear-flow or entropy-driven approaches, and therefore enables high alignment at low nanowire densities.

By integrating compressional and shear flow alignment within a single process, we further demonstrate programmable control over nanowire orientation, yielding micropatterned coatings with tunable anisotropy. The resulting films exhibit strong direction-dependent electrical, optical, and terahertz responses, and the approach is generalizable to a wide range of 1D nanomaterials. These findings establish a scalable, solution-based strategy for structuring anisotropic nanomaterials and pave the way for their incorporation into advanced optoelectronic, photonic, and sensing systems.

\section{Experimental Methods}
\subsection{Materials}

Silver nanowires (AgNWs) were purchased from XF Nano Co., Ltd.  
The nanowires had an average length of $60$~$\upmu$m and a diameter of 40~nm, and were suspended in deionized water at a concentration of 10~mg/mL.  
Carboxymethylcellulose sodium (CMC-Na, ${M}_{\text{w}} = 700{,}000$, $\text{DS} = 0.9$), ethanol, isopropyl alcohol, N-methylpyrrolidone, and n-octadecyl mercaptan were purchased from Macklin.  
All reagents were used as received without further purification.

\subsection{Preparation of AgNW suspensions containing CMC-Na}
CMC-Na was dissolved in deionized water under gentle stirring to prepare homogeneous solutions of 0.005~wt\%, 0.01~wt\%, 0.05~wt\%, and 0.1~wt\%.  
AgNW suspension was then added to each CMC-Na solution to obtain final AgNW concentrations of $0.25 \times 10^{-3}$~wt\%, $0.5 \times 10^{-3}$~wt\%, $1 \times 10^{-3}$~wt\%, $2 \times 10^{-3}$~wt\%, $3 \times 10^{-3}$~wt\%, $4 \times 10^{-3}$~wt\%, and $5 \times 10^{-3}$~wt\%.  
The resulting AgNW/CMC-Na suspensions were sonicated for 2 minutes at 300~W to ensure uniform nanowire dispersion.  
For comparison, AgNW suspensions at the same concentrations were also prepared in deionized water without CMC-Na as control samples.

\subsection{Optical imaging of AgNWs in suspensions and deposited coatings}

AgNWs in both suspensions and deposited coatings were imaged in reflection mode using a laser-scanning confocal microscope (Nikon Ti-E) equipped with a mercury light source.  
Confocal reflection microscopy (CRM) images of the deposited coatings were acquired using a 40$\times$/0.75 NA air objective.  
Time-lapse imaging was performed in a single focal plane over the substrate, with an exposure time of 5~ms, using a 20$\times$/0.5 NA air objective.

\subsection{Flow field measurements using microparticle image velocimetry ($\upmu$PIV)}

Yellow-green fluorescent tracer particles (0.11~$\upmu$m, carboxylate-modified, Molecular Probes) were added to Ag NW suspensions with varying concentrations of CMC-Na.  
The volume fraction of the tracer particles was $3 \times 10^{-3}$~vol\%.  
Time-lapse imaging during the drying process was performed using a fluorescent confocal microscope (Nikon Ti-E) equipped with a 20$\times$/0.5 NA air objective.  
The focal plane was set just above the substrate surface, and images were captured at 30~ms intervals.
$\upmu$PIV analysis was conducted using an open-source software package based on the MATLAB code by Thielicke and Stamhuis~\cite{stamhuis2014pivlab}.  
To reduce computational load, one of every five fluorescence images was used for velocity analysis, yielding an effective time step of 150~ms.  
Radial flow velocity was computed, and the radial coordinate $r$ was normalized by the initial droplet radius $R_0$ to facilitate comparison with theoretical predictions.

\subsection{Measurements of rheological property of CMC-Na solutions}

The local viscosity of CMC-Na aqueous solutions was measured using a micro-rheology approach.  
Yellow-green fluorescent tracer particles (0.11~$\upmu$m, carboxylate-modified, Molecular Probes) were added to CMC-Na solutions of varying concentrations at a volume fraction of $0.3 \times 10^{-3}$~vol\%.  
The positions and trajectories of the tracer particles were extracted from time-lapse optical images using publicly available MATLAB code~\cite{crocker1996methods, singleparticletrackingYX}.  
The mean square displacement (MSD) of each tracer was then analyzed using a standard micro-rheology algorithm~\cite{mason1995optical}.
To minimize the influence of outward convective flow, only the displacement components perpendicular to the radial direction (i.e., MSD$_\text{x}$) were used for viscosity calculation (see Supplementary Information for details).  
Each optical micrograph was segmented into eight radial regions, and the average local viscosity in each region was calculated from the MSD$_\text{x}$ values of the tracer particles located within that region.
Additionally, the bulk viscosities of CMC-Na aqueous solutions with different initial concentrations were measured using both a rotational viscometer (Brookfield DV3T) and the micro-rheology method for comparison.

\subsection{Characterization of electrical properties of AgNW coatings}

Sheet resistance measurements were performed using a Keithley 2450 digital source meter in conjunction with a JANDEL four-point probe setup.  
The probe consists of four thin metal needles arranged linearly with equal spacing of $d = 1.5$~mm.  
All measured samples were substantially larger than the probe array to ensure measurement accuracy.
A constant direct current (DC), $I$, was applied through the outer two probes, generating a voltage drop across the AgNW coating, which was detected by the inner two probes as $U$.  
The sheet resistance $R_s$ was calculated using the following expression:
\begin{equation}
    R_s = \frac{\pi}{\ln{2}} \frac{U}{I}
\end{equation}
To assess the electrical anisotropy of the AgNW coatings, the sample stage was rotated at an incremental angle of 15$^\circ$, and the sheet resistance was measured at various angles $\alpha$ relative to the nanowire alignment direction, all at the same spatial location on the sample.

\subsection{Characterization of optical properties of AgNW coatings}

Optical imaging of AgNW coatings was performed using a polarizing optical microscope (Nikon Eclipse Ci) equipped for both transmission and reflection modes.  
Samples were placed on the microscope stage and positioned between a polarizer and analyzer with orthogonal polarization axes.  
Incident light was directed normal to the sample surface.
To assess optical anisotropy, the AgNW coating was rotated through a range of angles relative to the alignment direction.  
At each angle, optical images were captured using a Nikon DS-Qi2 monochrome camera with a 4$\times$/0.13 NA air objective under identical imaging conditions.  
The grayscale intensity of each image was quantified by summing the pixel gray values, representing the total transmitted light intensity.
Color micrographs of the AgNW coatings were obtained in reflection mode using a QImaging Retiga R3 color camera and a 40$\times$/0.75 NA air objective.

\subsection{Characterization of terahertz polarization response of AgNW coatings}

The terahertz (THz) polarization response of the AgNW coatings was measured using an angle-resolved THz time-domain spectrometer operated in transmission mode.  
Linearly polarized THz pulses in the 0–1.5~THz frequency range were generated using a low-temperature-grown GaAs photoconductive antenna, excited by a femtosecond fiber laser oscillator (FemtoFErb FD 6.5, Toptica) with a central wavelength of 1560~nm and a repetition rate of 100~MHz.
The THz pulses were collimated using a 50~mm focal length lens and subsequently focused onto the AgNW coatings with a 100~mm lens.  
The transmitted waves were then collimated and collected using additional lenses with focal lengths of 100~mm and 50~mm, respectively.
A photoconductive antenna detector converted the transmitted THz signal into a photocurrent, which was sampled using a high-speed scanning delay line.  
The electric field vector of the THz wave was acquired through lock-in amplification.

\begin{acknowledgments}
The authors gratefully acknowledge Prof. Arjun G. Yodh for insightful discussions. 
We thank Prof. Sida Luo for providing the four-point probe setup for the sheet resistance measurements.
This work was supported by the National Natural Science Foundation of China (No. 12072010, No. 22473005, and No. 21961142020) and the Fundamental Research Funds for the Central Universities (YWF-22-K-101).
\end{acknowledgments}

\appendix

\section{Supporting Information}
\subsection{Preparation of free-standing flexible composite films with highly aligned silver nanowires (AgNWs)}

As an alternative viscosity-modifying polymer system, we used N-methylpyrrolidone (NMP) containing polyamic acid (PAA) to prepare free-standing AgNW/PAA composite films.  
Instead of dispersing AgNWs in an aqueous CMC-Na solution, AgNWs functionalized with octadecanethiol were dispersed in the PAA/NMP solution.  
The nanowire–polymer mixture was then deposited onto a glass substrate using the dip-coating evaporation setup shown in FIG.~1A of the main text.  
The resulting AgNW/PAA coating was easily detached by immersing the glass substrate in deionized water, yielding a free-standing flexible composite film, as shown in FIG.~S\ref{PI film}.  
The CRM image in the red dashed box reveals a high degree of nanowire alignment throughout the film, indicating directional organization over a large area during self-assembly.  
These results underscore the robustness of our approach for producing flexible composite films with uniformly aligned nanowire networks—an essential feature for devices requiring anisotropic optical or electronic performance.

\subsection{Micro-rheology measurements of CMC-Na solution}

The local viscosity of CMC-Na aqueous solutions was measured using the micro-rheology method.
Time-lapse imaging during the drying process of CMC-Na solutions containing tracer particles was performed using a fluorescent confocal microscope (Nikon Ti-E).
To quantify the distribution of the local viscosity within the droplet, each optical micrograph is equally divided into eight regions along the radius. 
The mean square displacement (MSD), $\left<\Delta d^2\right>$, was calculated by the fluorescent tracer particle trajectories from $x$ and $y$ coordinates
\begin{equation*}
    \left<\Delta d^2\right> = \left<(x(t+\tau)-x(t))^2\right>+\left<(y(t+\tau)-y(t))^2\right>.
\end{equation*}
Where $\tau=30$ ms is the lag time. The diffusion coefficient, $D$, was measured using
\begin{equation*}
    D=\left<\Delta d^2\right>/2nD\tau.
\end{equation*}
Where $n$ is the dimensionality. 
Note that, to avoid the effect of the evaporation-induced outward flow when measuring the local viscosity during drying, only the component of the displacements of tracer particles normal to the radial flow direction,
i.e. MSD$_\text{x}$ $(n=1)$, is used for measuring the local viscosity (FIG. S\ref{Micro-rheology}A-E).
When measuring the viscosity of the tranquil CMC-Na aqueous solution at 75$^\circ$C, displacements in both directions, i.e. MSD $(n=2)$, are utilized for calculation (FIG. S\ref{Micro-rheology}F).
Then, the viscosity $\eta$ can be calculated from the diffusion coefficient,
\begin{equation*}
   \eta = {k_B T}/{6\pi Dr}.
\end{equation*}
where $k_B$ is the Boltzmann constant, $T=348$ K is the absolute temperature, and $r=55$ nm is the radius of the fluorescent tracer particle. The local viscosity calculated from FIG. S\ref{Micro-rheology}A-E is plotted in FIG. S3C in main text, indicates a viscosity gradient along the radius inside the droplet during the drying process.

\subsection{Establishing the concentration–viscosity relationship in CMC-Na solutions}
To calculate the local concentration of CMC-Na from local viscosity, we characterized the relationship between the viscosity of CMC-Na aqueous solution and its concentration at 75$^\circ$C (FIG. S\ref{Viscosity to Concentration}).
The viscosity of CMC-Na aqueous solutions with concentrations ranging from 0 wt$\%$ to 0.1 wt$\%$ and 0.3 wt$\%$ to 1.0 wt$\%$ were measured by micro-rheology and viscometer, respectively (blue dots). 
The experimental measurement results achieve a good fit with previously reported theoretical calculations (red dashed line)\cite{CMCviscosity}:
\begin{align}
\eta_0 ={}& 0.891 + 7.82\times 10^{-3} w_0 \text{M}_\text{w}^{0.93} \notag \\
         & + 1.77 \times 10^{-5} w_0^2 \text{M}_\text{w}^{1.86} \notag \\
         & + 4.22 \times 10^{-12} w_0^{4.09} \text{M}_\text{w}^{3.8}
\end{align}
Here $\eta_0$ is the zero-shear viscosity of CMC-Na aqueous, $w_{0}$ is the concentration of CMC-Na aqueous, $\text{M}_\text{w}=700000$ is the weight-average molar mass of CMC-Na. Thus, we used the theoretical calculation equation to calculate the local concentrations, $w(r)$ from the local viscosity (Fig. 3C in main text).

\section{Supporting Information for Theoretical modeling and calculation of flow velocity pattern during evaporation }
\subsection{Basic model}\label{1.1}
\subsubsection*{Profile}
The CMC-Na aqueous solution can be considered as a binary component droplet containing non-volatile solutes. The droplet is placed on a substrate, size of which is less than the capillary length. Let $R(t)$ and $H(t)$ be the contact radius and the height at the center of the droplet, respectively. We assume that the contact angle is small ($H(t)/R(t)\ll 1$) the surface profile of the droplet at time, $t$, is given by a parabolic function as
\begin{equation}
    h(r,t) = H(t)\left(1 - \frac{r^2}{R^2}\right).
    \label{S_eq1}
\end{equation}
The contact angle, $\theta(t)$, is
\begin{equation}
    \theta(t) = \frac{2H(t)}{R(t)}.
    \label{S_eq2}
\end{equation}
The droplet volume, $V(t)$, is
\begin{equation}
    V(t) = \frac{\pi\theta(t)R^3(t)}{4}.
    \label{S_eq3}
\end{equation}
\subsubsection*{Evaporation}
In our theoretical calculation, water and CMC-Na are treated as two components of the solution, where water evaporation leads to the decrease in the volume, while CMC-Na is non-volatile. Therefore, the droplet volume, $V(t)$, changes with time, $t$, according to the following relationship
\begin{equation}
    \dot{V}(t) = -\int_{0}^{R(t)} 2\pi r J(r,t) \, dr,
    \label{S_eq4}
\end{equation}
where $J(r,t)$ is the evaporation flux, defined as the volume of liquid evaporating to air per unit time per unit surface area. Here, we assume that the solution is ideal, and use the following simple model for the evaporation rate
\begin{equation}
    J(r,t) = (1-c(r,t)) \cdot \frac{D_{\text{ev}}^W}{R(t)},
    \label{S_eq5}
\end{equation}
where $c(r,t)$ is the height-averaged volume concentration of CMC-Na in the droplet and ($1-c(r,t)$) is the height-averaged volume concentration of water. $D_{ev}^{W}$ is the diffusion coefficient of water to air and is influenced by environmental conditions such as temperature and relative humidity.
In this experiment, $D_{ev}^{W}$ is assumed to remain constant for different $c_0$ under the same temperature.

\subsubsection*{Conservation equation}
During evaporation, the conservation equation for the volume of liquid is written as
\begin{equation}
    \frac{\partial h}{\partial t} = -\frac{1}{r} \frac{\partial(rv_fh)}{\partial r} - J,
    \label{S_eq6}
\end{equation}
where $v_{f}(r,t)$ is the height-averaged velocity of the fluid at radius $r$ and time $t$. The mass conservation equation for water is written as
\begin{equation}
    \frac{\partial((1-c)h)}{\partial t} = -\frac{\partial(rv_f(1-c)h)}{\partial r}\frac{1}{r} - J,
    \label{S_eq7}
\end{equation}
where we have ignored the diffusion of each component. The first term on the right-hand side represents the effect of convection, and the second term represents the evaporation. Using Eq. (\ref{S_eq1}), we obtain the solve for $v_{f}(r,t)$ from Eq. (\ref{S_eq6}) expressed by $\dot{R}$ and $\dot{V}$
\begin{equation}
    v_f = \frac{r\dot{R}}{R} - \left(\frac{r}{2V} + \frac{r^3}{2\pi R^4 h}\right)\dot{V} - \frac{1}{rh} \int_{0}^{r} r' J(r',t) \, dr'.
    \label{S_eq8}
\end{equation}
Combining Eq. (\ref{S_eq6}) and (\ref{S_eq7}), we have the time derivative of $c(r,t)$ as
\begin{equation}
    \dot{c} = -v_f \frac{\partial c}{\partial r} + \frac{J}{h} c.
    \label{S_eq9}
\end{equation}
This equation indicates explicitly that even if the initial distribution of the CMC-Na is uniform $c(r,0)=c_0$, it will become non-uniform, which is caused by the evaporation effect, i.e., the second term in the right is always greater than 0 ($\dot{c}|_{t=0} = \frac{Jc_0}{h} > 0$).
\subsubsection*{Surface tension}
Here, since the initial concentration of CMC-Na is below 0.1 wt$\%$, it is allowed to ignore the influence of CMC-Na on the liquid/vapor surface tension of the solution. And we take advantage of the Young equation
\begin{equation}
    \theta_e = \sqrt{\frac{2(\gamma_{\text{LS}} + \gamma_{\text{LV}} - \gamma_{\text{SV}})}{\gamma_{\text{LV}}}},
    \label{S_eq10}
\end{equation}
where $\theta_e$ is the equilibrium contact angle. Here, $\gamma_{\text{LV}}$, $\gamma_{\text{LS}}$ and $\gamma_{\text{SV}}$ are the liquid/vapor, liquid/substrate, and substrate/vapor interface tension of water, respectively. In the calculation, we take use of the small contact angle assumption. Then the approximation ($1-\theta_{e}^{2}/2$) replaces $cos\theta_e$ in $cos\theta_e=(\gamma_{\text{SV}}-\gamma_{\text{LS}})/\gamma_{\text{LV}}$.
\subsection{Rayleigh function}
The time evolution equation of $\dot{R}(t)$ is derived by taking the variation of the Rayleigh of the system based on Onsager principle
\begin{equation}
    \mathcal{R} = \Phi + \dot{F}.
    \label{S_eq11}
\end{equation}
Here, $\Phi$ is the energy dissipation function and $\dot{F}$ is the time change of the free energy of the system. The specific form of the two parts is presented below.
\subsubsection*{The free energy}
The free energy $F$ is the sum of the interfacial energy and can be written as
\begin{equation}
    F = \int_{0}^{R} \gamma_{\text{LV}} 2\pi r \sqrt{1 + h'^2} \, dr + (\gamma_{\text{LS}} - \gamma_{\text{SV}})\pi R^2.
    \label{S_eq12}
\end{equation}
then the time derivative of the free energy is easily calculated as
\begin{equation}
    \dot{F} = \gamma_{\text{LV}} \left[\left( -\frac{16V^2}{\pi R^5} + \pi \theta_e^2 R \right)\dot{R} + \frac{8V\dot{V}}{\pi R^4}\right].
    \label{S_eq13}
\end{equation}
\subsubsection*{The Energy Dissipation Function}
According to the experimental results of the evaporation rate of the CMC-Na aqueous, higher initial concentration of CMC-Na, $c_0$, causes much greater viscosity of the solution, $\eta$. Therefore, the energy dissipation in the present problem includes two parts, which are the shear dissipation within the CMC-Na aqueous as well as the friction between the contact line and the substrate. Then, $\Phi$ is written as
\begin{equation}
    \Phi = \frac{1}{2} \int_{0}^{R} \int_{0}^{2\pi} \frac{3\eta}{h} v_f^2 r \, dr \, d\alpha + \pi \xi_{\text{cl}} R \dot{R}^2,
    \label{S_eq14}
\end{equation}
where $\xi_{\text{cl}}$ is a phenomenological parameter representing the mobility of the contact line. Inserting Eq. (\ref{S_eq8}) into Eq. (\ref{S_eq14}), we have
\begin{align}
\Phi ={} & \, 3\pi\eta \int_{0}^{R} \frac{1}{h} 
\Bigg( r\frac{\dot{R}}{R} 
- \left( \frac{r}{2V} + \frac{r^3}{2\pi R^4 h} \right)\dot{V} \notag \\
&\quad - \frac{1}{r h} \int_{0}^{r} r' J(r',t) \, dr' 
\Bigg)^2 r \, dr 
+ \pi \xi_{\text{cl}} R \dot{R}^2.
\label{S_eq15}
\end{align}
\subsection{Evolution function of $\Dot{R}$}
The time evolution of $\dot{R}(t)$ is determined by the condition ${\partial R}/{\partial \dot{R}} = 0$. Here, we calculate ${\partial R}/{\partial \dot{R}}$ by the sum of ${\partial \dot{F}}/{\partial \dot{R}}$ and ${\partial \Phi}/{\partial \dot{R}}$
\begin{equation}
\begin{aligned}
    \frac{\partial R}{\partial \dot{R}} &= \pi R \gamma_{\text{LV}} (\theta_e^2 - \theta^2) + \frac{3\pi^2 \eta \alpha R^4}{2V} \dot{R} + \frac{3\pi^2 \eta R^5 \dot{V}}{8V^2} \\
    & + 6\pi \eta \int_{0}^{R} \frac{r}{h^2 R} \left( \int_{r}^{R} r' J(r') \, dr' \right) \, dr + 2\pi \xi_{\text{cl}} R \dot{R} = 0.
    \label{S_eq16} 
\end{aligned}
\end{equation}
Then we get the $\dot{R}$
\begin{align}
&\left( \frac{3\pi \eta \alpha R^3}{2V} + 2\xi_{\text{cl}} \right) \dot{R} 
= \gamma_{\text{LV}} (\theta^2 - \theta_e^2) 
- \frac{3\pi \eta R^4 \dot{V}}{8V^2} \notag \\
&\quad - \frac{6\eta}{R^2} \int_{0}^{R} \frac{r}{h^2} 
\left( \int_{r}^{R} r' J(r') \, dr' \right) \, dr.
\label{S_eq17}
\end{align}
\subsection{Scale for the evolution function}
To unify the standards, we conduct the scaling by two characteristic time and two dimensionless parameters. We define the characteristic time of evaporation $\tau_{\text{ev}} = -\frac{V_0}{\dot{V}_0}$, where $V_0 = \frac{\pi}{4} \theta_0 R_0^3$ and $\dot{V}_0 = -\pi R_0 D_{\text{ev}}^W$. Then $\tau_{\text{ev}}$ is written as
\begin{equation}
    \tau_{\text{ev}} = \frac{\theta_0 R_0^2}{4D_{\text{ev}}^W}.
    \label{S_eq18}
\end{equation}
We define the characteristic relaxation time of the droplet determined by viscosity $\eta$ and the surface tension $\gamma_{LV}$.
\begin{equation}
    \tau_{\text{re}} = \frac{\eta V_0^{\frac{1}{3}}}{\gamma_{\text{LV}}}.
    \label{S_eq19}
\end{equation}
Besides, we introduce two key dimensionless parameters, which determine the evolution process. The first one represents the evaporation rate, written as
\begin{equation}
    k_{\text{ev}} = \frac{\tau_{\text{re}}}{\tau_{\text{ev}}} = \frac{4\eta D_{\text{ev}}^W V_0^{\frac{1}{3}}}{\gamma_{\text{LV}} \theta_0 R_0^2}.
    \label{S_eq20}
\end{equation}
The second one is used to characterize the strength of the friction effect, written as
\begin{equation}
    k_{\text{cl}} = \frac{\xi_{\text{cl}}}{\xi_{\text{hydro}}} = \frac{\theta \xi_{\text{cl}}}{3\alpha\eta}.
    \label{S_eq21}
\end{equation}
Here, $\xi_{\text{hydro}} = {3\alpha\eta}/{\theta}$ denotes the friction constant of the solution calculated by hydrodynamics. Combining Eq. (\ref{S_eq18})-(\ref{S_eq21}), Eq. (\ref{S_eq17}) can be reduced to
\begin{align}
(1 + k_{\text{cl}}) \dot{R} ={}& \frac{V_0^{1/3} \theta (\theta^2 - \theta_e^2)}{6\alpha \tau_{\text{re}}}
- \frac{R \dot{V}}{4\alpha V} \notag \\
& - \frac{\theta \theta_0 R_0^2}{4\alpha R^3 \tau_{\text{ev}}}
\int_{0}^{R} \frac{r}{h^2}
\left( \int_{r}^{R} r' (1 - c(r')) \, dr' \right) dr.
\label{S_eq22}
\end{align}
\subsection{Numerical calculation}
In the numerical calculation, we rewrite Eq. (\ref{S_eq22}) as
\begin{align}
(1 + k_{\text{cl}}) \tau_{\text{ev}} \dot{R} ={}& 
\frac{V_0^{1/3} \theta (\theta^2 - \theta_e^2)}{6\alpha k_{\text{ev}}}
- \frac{\tau_{\text{ev}} R \dot{V}}{4\alpha V} \notag \\
& - \frac{\theta \theta_0 R_0^2}{4\alpha R^3}
\int_{0}^{R} \frac{r}{h^2}
\left( \int_{r}^{R} r' (1 - c(r')) \, dr' \right) dr.
\label{S_eq23}
\end{align}
The velocity can be rewritten as
\begin{align}
\tau_{\text{ev}} v_f ={}& \, \tau_{\text{ev}} r \frac{\dot{R}}{R}
- \left( \frac{r}{2V} + \frac{r^3}{2\pi R^4 h} \right) \tau_{\text{ev}} \dot{V} \notag \\
& - \frac{\theta_0 R_0^2}{4 r h R}
\int_{0}^{r} r' (1 - c(r')) \, dr'.
\label{S_eq24}
\end{align}
The volume change rate can be rewritten as
\begin{align}
\tau_{\text{ev}} \dot{V}(t) 
={}& -\tau_{\text{ev}} \int_{0}^{R(t)} 2\pi r J(r,t) \, dr \notag \\
={}& -\frac{\pi \theta_0 R_0^2}{2R} \int_{0}^{R} r (1 - c) \, dr.
\label{S_eq25}
\end{align}
The time change rate of $c(r,t)$ can be rewritten as
\begin{equation}
    \tau_{\text{ev}} \dot{c} = -\tau_{\text{ev}} v_f \frac{\partial c}{\partial r} + \frac{\theta_0 R_0^2}{4Rh} c(1-c).
    \label{S_eq26}
\end{equation}

In summary, the initial conditions include $k_{\text{ev}}$, $k_{\text{cl}}$, $c_0$, $\theta_0$, $\theta_e$. We set $R_0 = 1.0$, which facilitates the contrast and avoids the influence of different initial values in experiments. After the droplet is placed on the substrate, the contact angle quickly reaches the equilibrium state. Therefore, we set $\theta_0 = \theta_e = 0.45$ in the calculation according to the measurement (Fig. S\ref{Contact angle}). The parameter $k_{\text{ev}}$ is set as constant $k_{\text{ev}} = 0.02$. Since the contact line is pinned during most of the evaporation process, $k_{\text{cl}}$ is set to $k_{\text{cl}} = 25$ to mimic the large friction

Suppose we know $R(t)$ and the volume fraction distribution $c(r,t)$ at time $t$, $\dot{V}(t)$ is calculated by Eq. (\ref{S_eq25}), $\theta(t)$ is given by droplet volume function. Moreover, the velocity $v_f(r,t)$ is expressed as a function of $\dot{R}$, $\dot{V}$ and $c(r,t)$ by Eq. (\ref{S_eq24}), and the distribution evolution $c(r,t)$ is a function of $v_f(r,t)$ in Eq. (\ref{S_eq26}). Therefore, the time evolution of $V(t)$, $R(t)$ and $\theta(t)$ can be calculated, if $\dot{R}$ is known. And $\dot{R}$ is determined by the Onsager principle shown in Eq. (\ref{S_eq23}).

\subsection{Discussion of the velocity}
Since the contact line is pinned during most of the process, i.e., $\dot{R}=0$, the velocity in Eq. (\ref{S_eq8}) can be reduced to,
\begin{equation}
    v_f = -\left(\frac{r}{2V} + \frac{r^3}{2\pi R^4 h}\right) \dot{V} - \frac{1}{rh} \int_{0}^{r} r' J(r',t) \, dr'.
    \label{S_eq27}
\end{equation}
When there is no CMC-Na in the solution, i.e., $c_0=0$, the evaporation flux has the form $J(r,t)={D_{\text{ev}}^W}/{R(t)}$. And the velocity can be rewritten as
\begin{align}
v_f ={}& \, 2\pi \left( \frac{r}{2V} + \frac{r^3}{2\pi R^4 h} \right)
\int_{0}^{R(t)} \frac{r D_{\text{ev}}^W}{R} \, dr 
- \frac{D_{\text{ev}}^W r}{2 h R} \notag \\
={}& \, \frac{\pi D_{\text{ev}}^W r R}{4V}.
\label{S_eq29}
\end{align}
which illustrates that $v_f$ is proportional to $r$, shown as a straight line in FIG 3D in main text. When CMC-Na is added to the solution, that is $c_0 \neq 0$, the evaporation flux has the form $J(r,t) = (1-c(r,t)) \frac{D_{\text{ev}}^W}{R(t)}$. And the velocity can be rewritten as
\begin{align}
v_f ={}& \, 2\pi D_{\text{ev}}^W \left( \frac{r}{2V} + \frac{r^3}{2\pi R^4 h} \right)
\int_{0}^{R(t)} \frac{r(1 - c(r,t))}{R(t)} \, dr \notag \\
& - \frac{D_{\text{ev}}^W}{r h} \int_{0}^{r} \frac{r'(1 - c(r',t))}{R(t)} \, dr' \notag \\
={}& \, \frac{\pi D_{\text{ev}}^W r R}{4V}
- \frac{2\pi D_{\text{ev}}^W}{R} \left( \frac{r}{2V} + \frac{r^3}{2\pi R^4 h} \right)
\int_{0}^{R(t)} r c(r,t) \, dr \notag \\
& + \frac{D_{\text{ev}}^W}{r R h} \int_{0}^{r} r' c(r',t) \, dr'.
\label{S_eq30}
\end{align}

To facilitate the discussion, we define the three terms in the right as 
\begin{align*}
    &\Gamma_1 = \frac{\pi D_{\text{ev}}^W rR}{4V},\\
    &\Gamma_2 = -\frac{2\pi D_{\text{ev}}^W}{R} \left( \frac{r}{2V} + \frac{r^3}{2\pi R^4 h} \right) \int_{0}^{R(t)} rc(r,t) \, dr,\\
    &\Gamma_3 = \frac{D_{\text{ev}}^W}{rRh} \int_{0}^{r} r' c(r',t) \, dr' \text{, respectively}.
\end{align*} It is obvious that $\Gamma_1$ is just the result of $v_f$ when $c_0=0$. $\Gamma_2$ ($<0$) and $\Gamma_3$ ($>0$) are related to the distribution of $c(r,t)$. The joint action of $\Gamma_2$ and $\Gamma_3$ makes $v_f$ no longer a linear function of $r$. 

\section{Dynamics of fibers in fluid flows with velocity and viscosity gradients}
Consider a rigid fiber of length $L=2a$ moving in a viscous fluid with a viscosity gradient. Denote the signed arc length as $s\in [-a, a]$. We assume that the fiber has a tapered shape with the cross-sectional radius $\lambda(s) = b\sqrt{1-s^2/a^2}$. The fiber aspect ratio $\epsilon = b/a \ll 1$. We further assume that the fiber is short compared with the characteristic spatial scales of variations in both fluid flow and viscosity. In current experiments, the viscosity gradient aligns the flow direction $\hat{\mathbf{r}}$. The fiber centerline is described as $\mathbf{x}(s, t) = \mathbf{x}_c + s\mathbf{p}$, where $\mathbf{x}_c$ is the center-of-mass position and the unit tangent vector of the fiber $\mathbf{p} = (\sin\varphi, \cos\varphi)$. The fiber velocity 
\begin{equation}\label{fiber_velo}
\mathbf{u} = \frac{d\mathbf{x}}{dt} = \dot{\mathbf{x}}_c + s\dot{\mathbf{p}} = \dot{\mathbf{x}}_c - s\mathbf{p}^{\perp}\dot{\varphi},
\end{equation}
where the unit normal vector $\mathbf{p}^{\perp} = (-\cos\varphi, \sin\varphi)$. We write the viscosity and the flow velocity along the fiber centerline as
\begin{gather}
\eta(s) = \eta_0 (\mathbf{x}_c) \left[1 + \frac{\tilde{\kappa}_\eta}{2a}s (\hat{\mathbf{r}}\cdot\mathbf{p})\right],\label{vis} \\
\mathbf{v}(s) = v_0 (\mathbf{x}_c) \hat{\mathbf{r}} \left[1 + \frac{\tilde{\kappa}_v}{2a} s (\hat{\mathbf{r}}\cdot\mathbf{p})\right],\label{flow_velo}
\end{gather}
where $\eta_0 (\mathbf{x}_c)$ and $v_0 (\mathbf{x}_c)$ are the reference viscosity and velocity at the center of the fiber, respectively, and $\tilde{\kappa}_\eta$ and $\tilde{\kappa}_v$ are the nondimensionalized gradients, 
\begin{equation}
\tilde{k}_{\eta} = \frac{2a}{\eta_0 (\mathbf{x}_c)} \kappa_{\eta},\quad \tilde{\kappa}_v = \frac{2a}{v_0 (\mathbf{x}_c)} \kappa_v.
\end{equation}
We nondimensionalize the system by scaling lengths by $a$ (radius by $b$), viscosity by $\eta_0$, and velocity by $v_0$, and denote the nondimensional quantities with $\tilde{\ }$:
\begin{align}
\tilde{s} &= \frac{s}{a}, \quad
\tilde{\lambda} = \frac{\lambda}{b}, \quad
\tilde{\mathbf{x}} = \frac{\mathbf{x}}{a}, \notag \\
\tilde{\eta} &= \frac{\eta}{\eta_0}, \quad
\tilde{\mathbf{u}} = \frac{\mathbf{u}}{v_0}, \quad
\tilde{\mathbf{v}} = \frac{\mathbf{v}}{v_0}.
\label{nondim_vars}
\end{align}

Based on a resistive-force theory for slender bodies in viscosity gradients\cite{Kamal_Lauga_2023}, the hydrodynamic force density $\tilde{\mathbf{f}}$ on the fiber is given by
\begin{align}
\tilde{\mathbf{f}}(s) ={}& \,
2\pi\tilde{\eta} \left[
\frac{\tilde{\mathbf{v}} - \tilde{\mathbf{u}}}{\ln \epsilon}
+ \frac{
\mathbf{J} + (\tilde{\mathbf{v}} - \tilde{\mathbf{u}})\ln(2\epsilon/\tilde{\lambda})
}{(\ln \epsilon)^2}
\right] \cdot
\left[
\frac{d\tilde{\mathbf{x}}}{d\tilde{s}} \frac{d\tilde{\mathbf{x}}}{d\tilde{s}} - 2\mathbf{I}
\right] \notag \\
& +
\frac{\tilde{\mathbf{v}} - \tilde{\mathbf{u}}}{2(\ln \epsilon)^2} \cdot
\left[
3\frac{d\tilde{\mathbf{x}}}{d\tilde{s}} \frac{d\tilde{\mathbf{x}}}{d\tilde{s}} - 2\mathbf{I}
\right].
\label{force}
\end{align}
Here, the integral $\mathbf{J}$ is defined as
\begin{align}
J_i = \lim_{\epsilon \to 0} \frac{\tilde{\eta}}{2} \Biggl(
\int_{-1}^{s-\epsilon} + \int_{s+\epsilon}^{1} \Biggr)
& \tilde{G}_{ij} \big[(\tilde{s} - \tilde{s}') \mathbf{p}\big]
\left[
\delta_{jk} - \frac{1}{2} \left(
\frac{d\tilde{\mathbf{x}}}{d\tilde{s}} \frac{d\tilde{\mathbf{x}}}{d\tilde{s}}
\right)\Big|_{\tilde{s} = \tilde{s}'}
\right] \notag \\
& \times
\big[\tilde{\mathbf{v}}(\tilde{s}') - \tilde{\mathbf{u}}(\tilde{s}')\big]
\, d\tilde{s}'.
\label{J}
\end{align}
where
\begin{align}
\tilde{G}_{ij}(\tilde{\mathbf{X}}) = \frac{1}{\tilde{\eta}} \Biggl[
& \left(
\frac{\delta_{ij}}{\tilde{X}} + \frac{\tilde{X}_i \tilde{X}_j}{\tilde{X}^3}
\right) \left(1 + \frac{\tilde{\kappa}_1 \tilde{X}_1}{4}\right) \notag \\
& + \frac{1}{4} \frac{
\tilde{X}_i \tilde{\kappa}_j \delta_{1j} - \tilde{\kappa}_i \tilde{X}_j \delta_{1i}
}{\tilde{X}}
\Biggr].
\end{align}
with $\tilde{\kappa}_1 = \tilde{\kappa}_\eta (\hat{\mathbf{r}}\cdot\mathbf{p})$ the viscosity gradient along the fiber orientation.

Based on Eqs~(\ref{fiber_velo})--(\ref{flow_velo}), (\ref{force}), and (\ref{J}), we can compute the force density $\tilde{\mathbf{f}}$ acting on the fiber. Since the fiber is passively advected by the fluid flow, the total hydrodynamic force and torque are both zero,
\begin{equation}
\int_{-1}^{1}\tilde{\mathbf{f}}\,d\tilde{s} = 0,\quad \int_{-1}^{1}\tilde{s}\mathbf{p}\times\tilde{\mathbf{f}}\,d\tilde{s} = 0.
\end{equation}
From these two constants, we obtain the center-of-mass velocity and rotational velocity,
\begin{gather}
\tilde{\mathbf{u}} \cdot \mathbf{p} = \sin\varphi + \mathcal{O}(\tilde{\kappa}_\eta \tilde{\kappa}_v), \\
\tilde{\mathbf{u}} \cdot \mathbf{p}^{\perp} = -\cos\varphi, \\
\dot{\varphi} = \frac{1}{2}\tilde{\kappa}_v \sin\varphi \cos\varphi.
\end{gather}
Restoring dimensions,
\begin{equation}
\dot{\varphi} = \frac{dv}{dr} \sin\varphi\cos\varphi.
\end{equation}

\newpage
\onecolumngrid
\section{Supplementary Figures}
\setcounter{figure}{0}
\captionsetup[figure]{labelformat=nospace, labelsep=period}

\begin{figure*}[h!]
\centering
\includegraphics[width=0.45\textwidth]{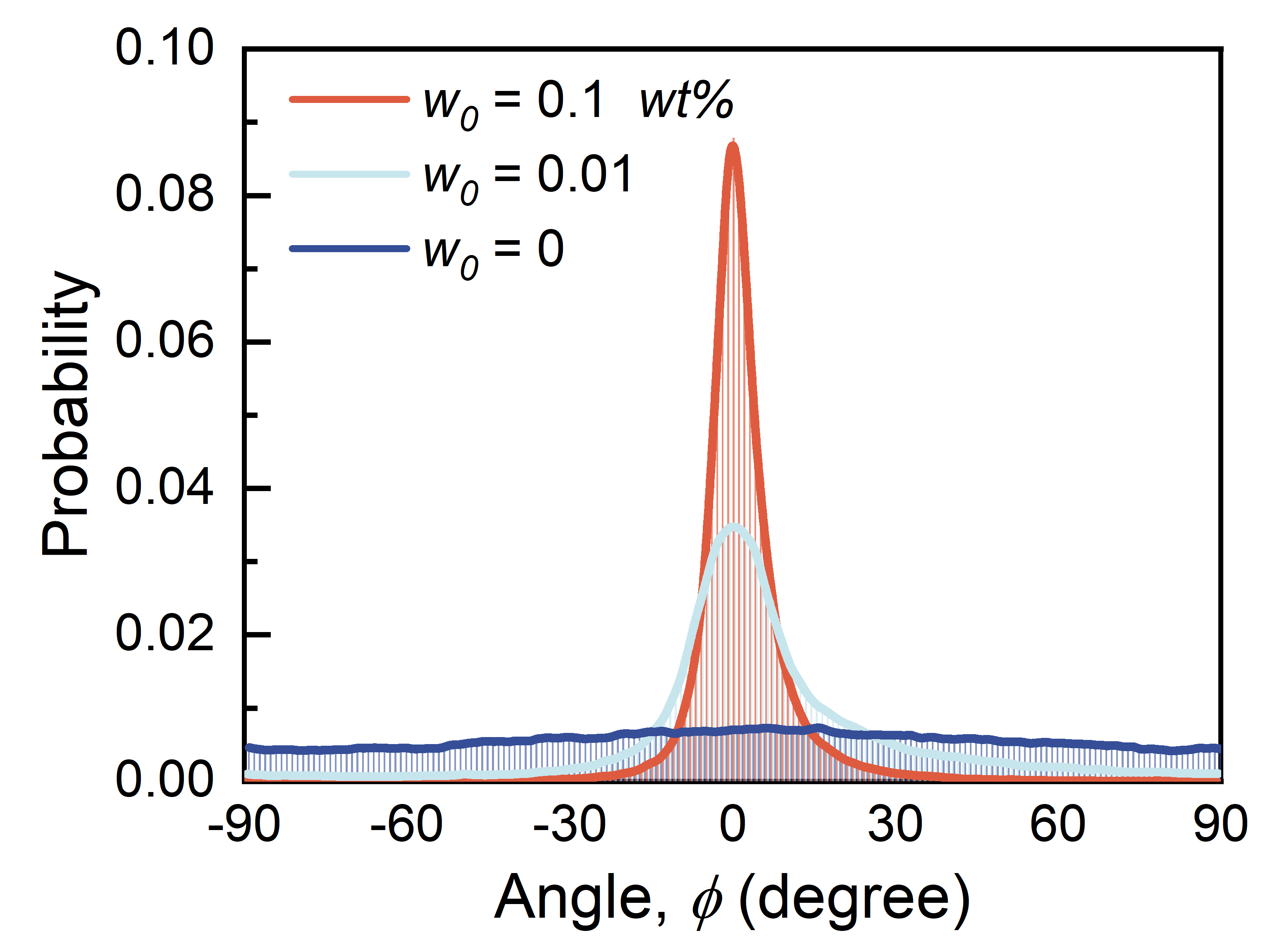}
\caption{Histograms showing the distribution of angles between the orientation of nanowires and a predefined reference direction across three representative samples.}
\label{Probability}
\end{figure*}

\begin{figure}[h!]
\centering
\includegraphics[width=0.45\textwidth]{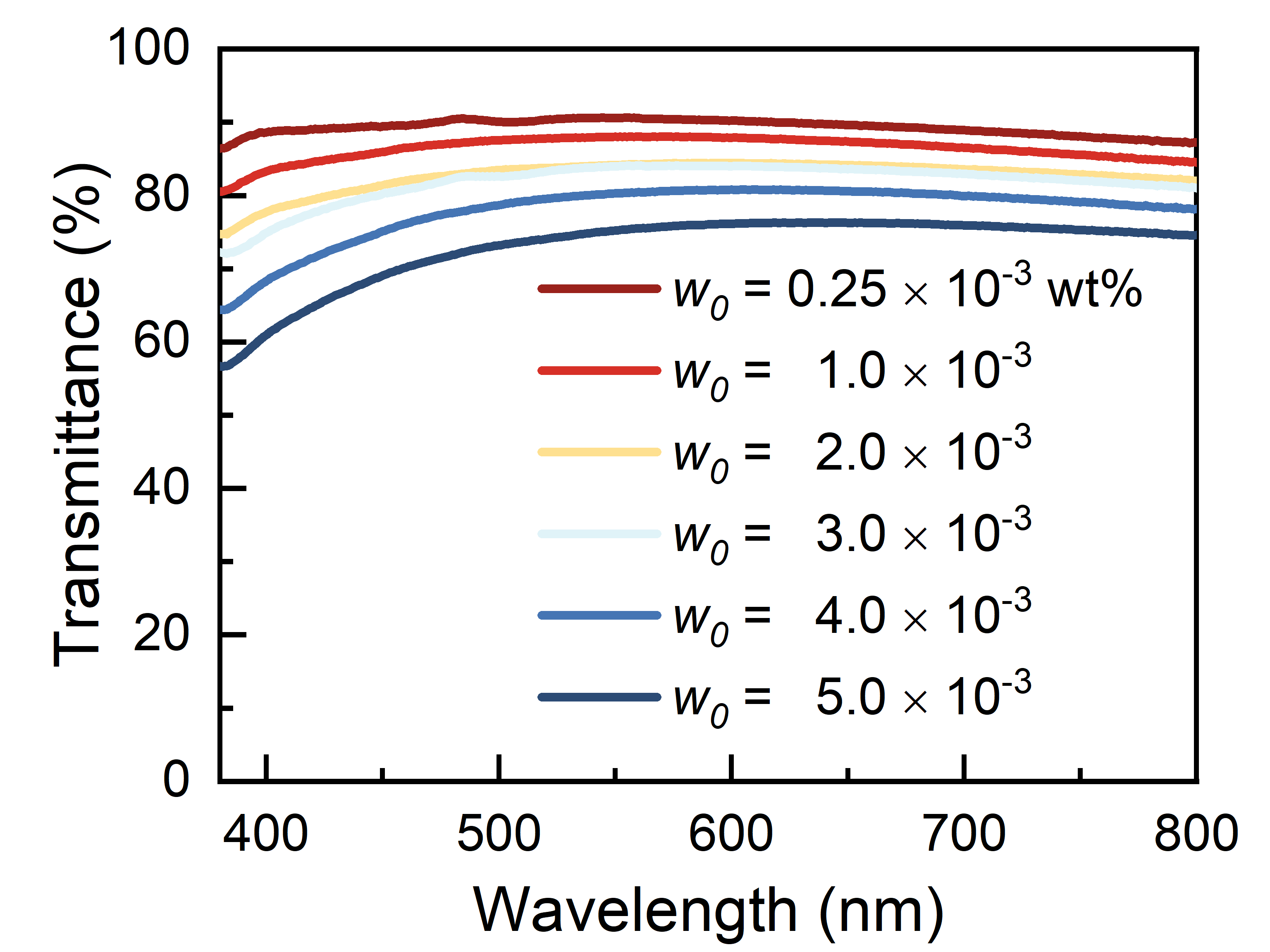}
\caption{Spectra of transmitted light of AgNW coatings prepared with different initial concentrations of AgNWs.}
\label{Transmittance}
\end{figure}

\begin{figure}[h!]
\centering
\includegraphics[width=0.6\textwidth]{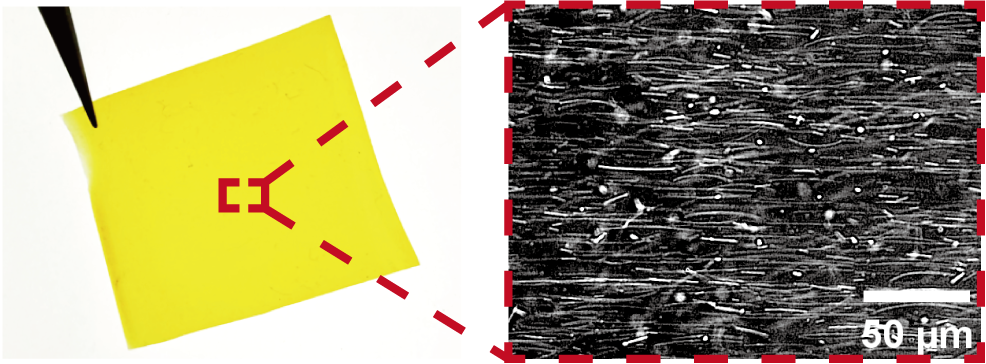}
\caption{Image of polyimide composite films with ordered AgNWs. In the red dashed box is the CRM image of the ordered AgNWs in the film} 
\label{PI film}
\end{figure}

\begin{figure*}[h!]
\centering
\includegraphics[width=0.3\textwidth]{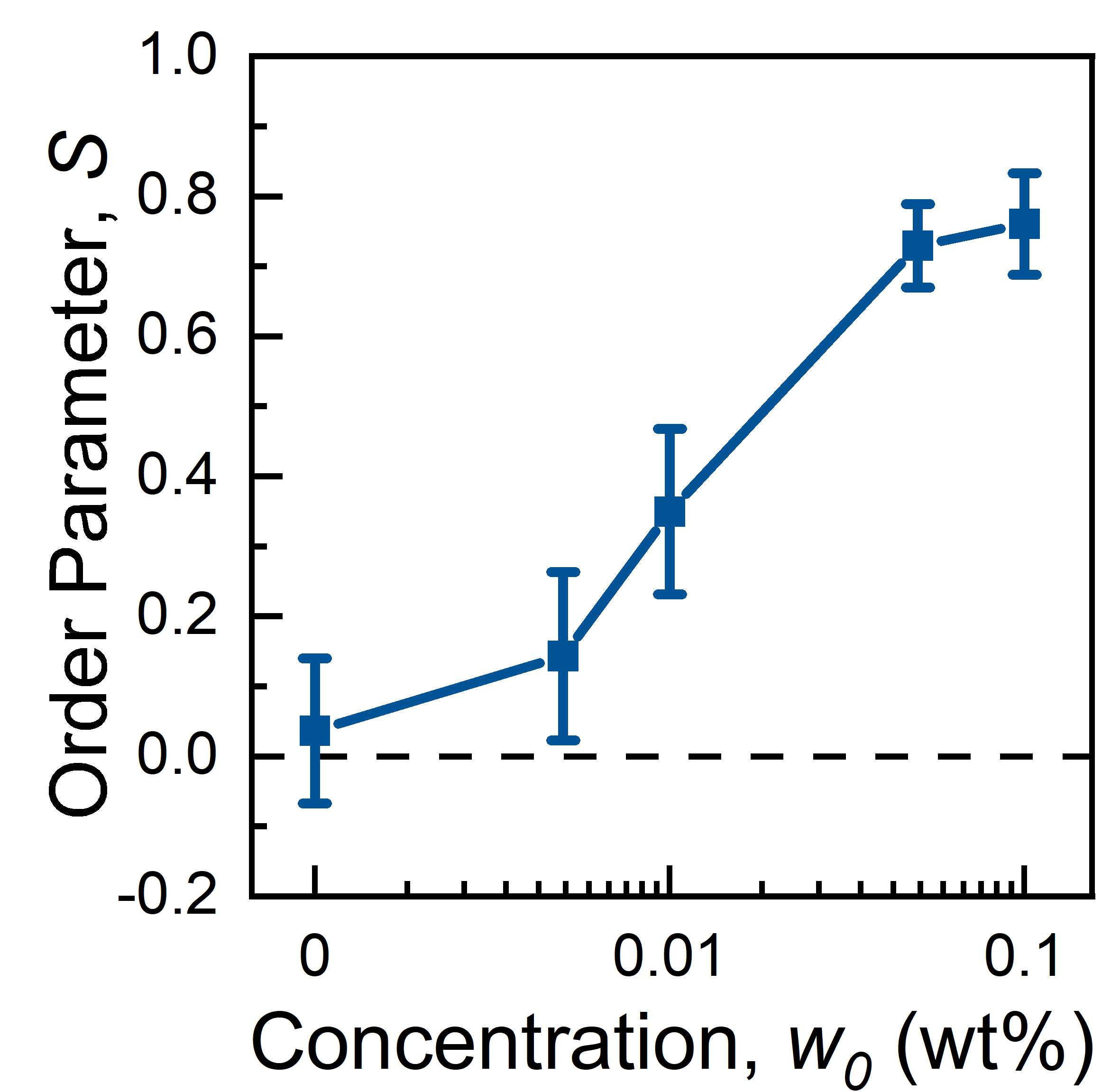}
\caption{The order parameter, $S$, calculated for AgNW coatings prepared under droplet evaporation setup, plotted against the initial CMC-Na concentration ($w_0$) in the suspensions.}
\label{Order Parameter-Droplet}
\end{figure*}

\begin{figure*}[h!]
\centering
\includegraphics[width=0.85\textwidth]{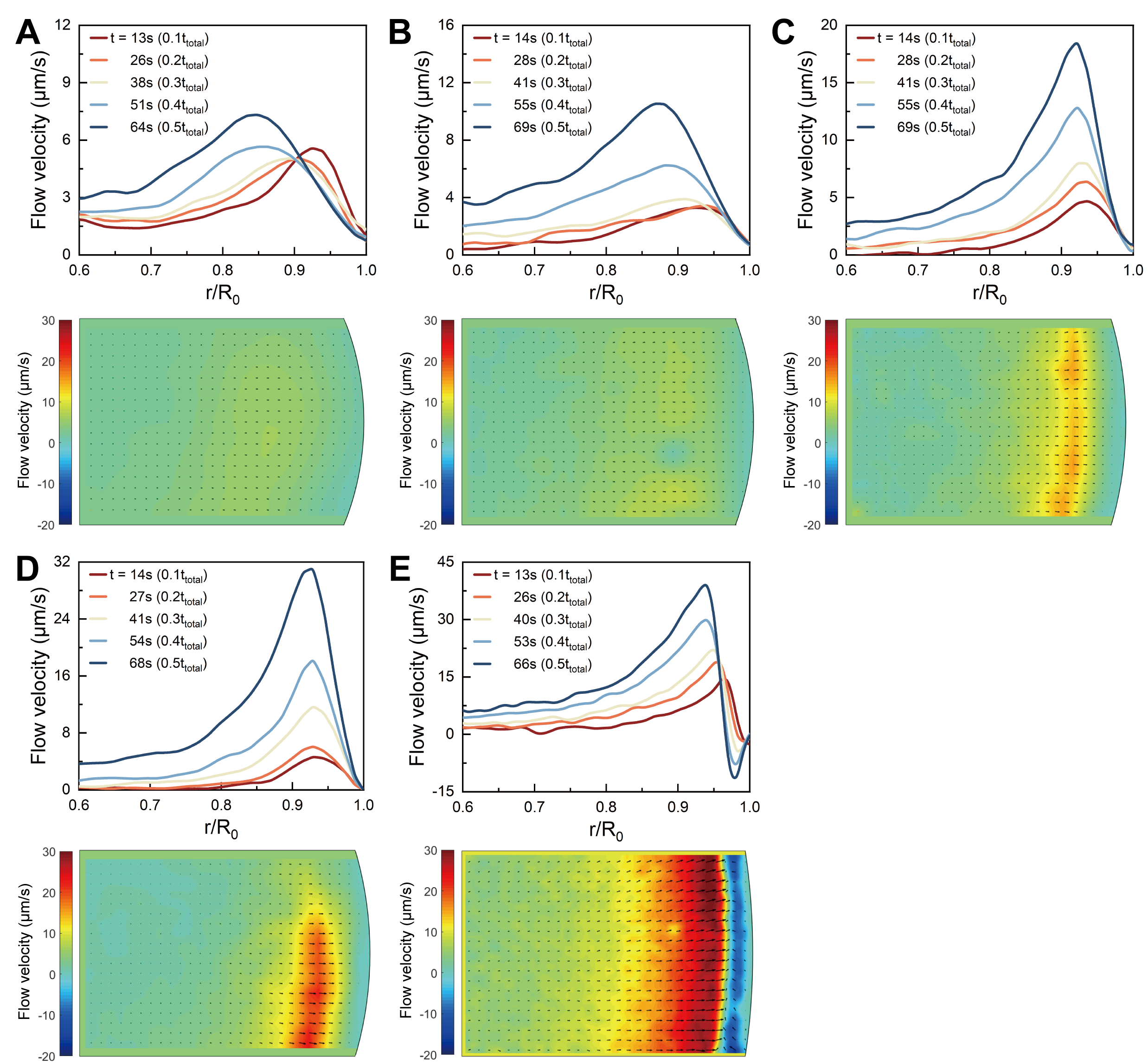}
\caption{Measurement of the flow velocity in the drying process by $\upmu$PIV. 
(A)-(E) Flow velocity, $v$, versus normalized radius, $r/R_0$, at different times, alongside the color map illustrating the flow velocity field when $t=0.4_\text{total}$. The initial concentrations of CMC-Na in droplets are (A) 0.1 wt$\%$, (B) 0.05 wt$\%$, (C) 0.01 wt$\%$, (D) 0.005 wt$\%$, and (E) 0 wt$\%$.}
\label{PIV}
\end{figure*}

\begin{figure*}[h!]
\centering
\includegraphics[width=0.45\textwidth]{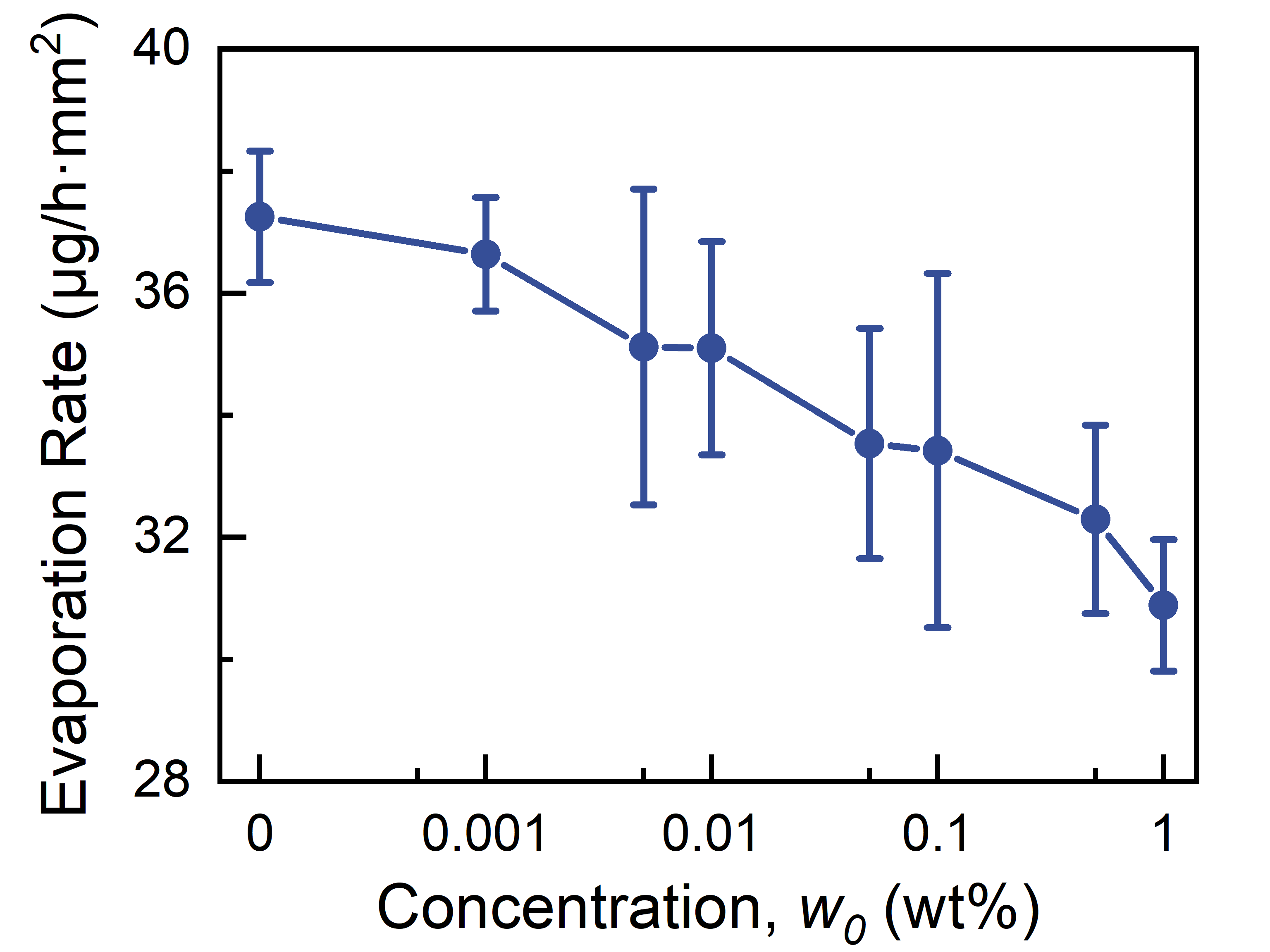}
\caption{The evaporation rate of the CMC-Na aqueous solution with the initial Concentrations ranging from 0 wt\textit{\%} to 1.0 wt\textit{\%}.
The evaporation rate (liquid mass evaporating to air per unit time per unit surface area) is measured under ambient temperature ($21\pm1^\circ$C) and relative humidity of $30\pm5\%$.
5 mL of CMC-Na aqueous solution is placed in a circular tube with a diameter $D=7$ mm, and the evaporation rate is calculated by measuring the change in liquid mass over the same time (168 hours). The measured results indicate that as the concentration of CMC-Na increases, the evaporation rate of the CMC-Na aqueous solution decreases.}
\label{Evaporation Rate}
\end{figure*}

\begin{figure*}[h!]
\centering
\includegraphics[width=0.45\textwidth]{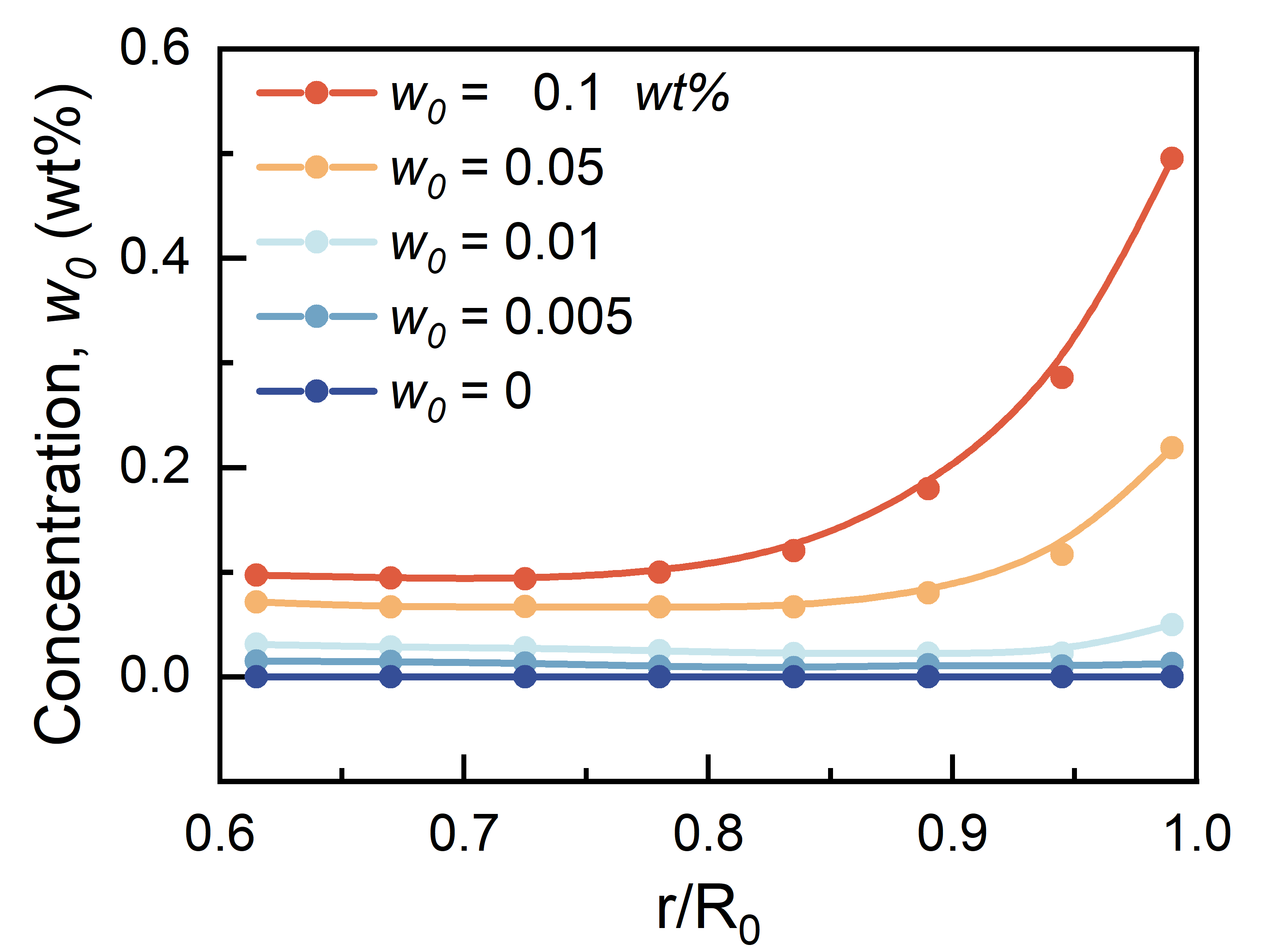}
\caption{Local concentration of the CMC-Na calculated by the local viscosity as a function of normalized radius, r/R0, inside CMC-Na aqueous droplets with different initial concentrations.}
\label{Concentrations distribuion}
\end{figure*}

\begin{figure*}[h!]
\centering
\includegraphics[width=0.5\textwidth]{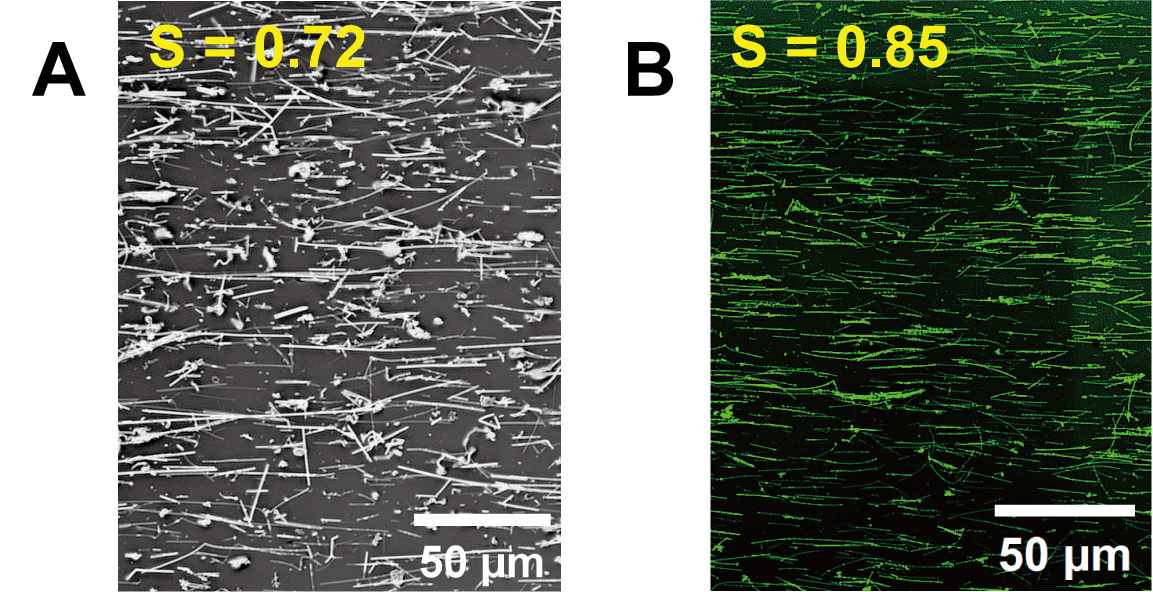}
\caption{Fabrication of aligned nanowire coatings with different types of nanowires. 
(\textit{A}) CRM image of the aligned SiC nanowire coating, with order parameter $S=0.72$. 
(\textit{B}) Fluorescent micrograph of the aligned SiO$_2$ nanowire coating, with order parameter $S=0.85$.}
\label{Different NWs}
\end{figure*}

\begin{figure*}[h!]
\centering
\includegraphics[width=0.6\textwidth]{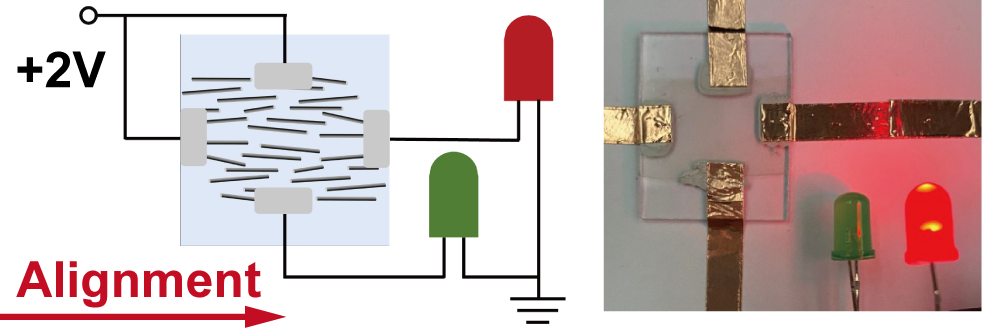}
\caption{Schematic and photograph of the aligned coating with $S=0.9$ unidirectional conduction along the alignment direction, enabling selective illumination of LEDs.}
\label{Electric anisotropy-illumination of LEDs}
\end{figure*}

\begin{figure*}[h!]
\centering
\includegraphics[width=0.6\textwidth]{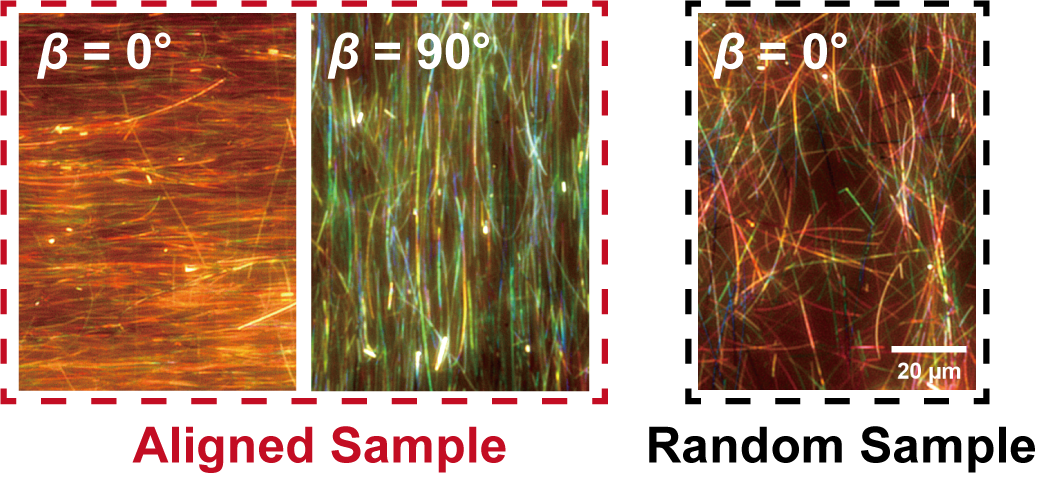}
\caption{Optical micrographs of the AgNW coating imaged with the reﬂection mode. Micrographs inside the red dashed box are for the well-aligned coating with $S=0.91$ and the one inside the black dashed box is for the coating of randomly-oriented AgNWs ($S=0.02$).}
\label{Combined with Dip-coating}
\end{figure*}

\begin{figure*}[h!]
\centering
\includegraphics[width=0.45\textwidth]{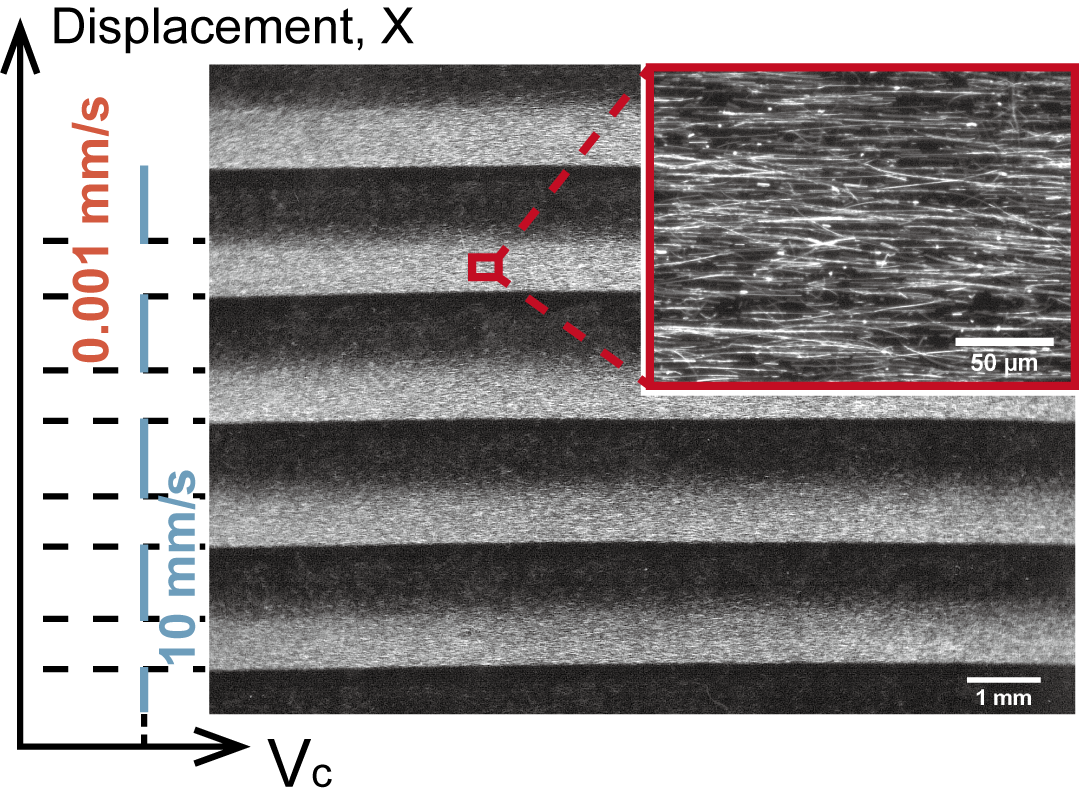}
\caption{CRM images of AgNW coatings prepared at fast withdrawal output modes ($v_{c}=10$ mm/s) of nanomotor.}
\label{Combined with Dip-coating}
\end{figure*}

\begin{figure*}[h!]
\centering
\includegraphics[width=0.95\textwidth]{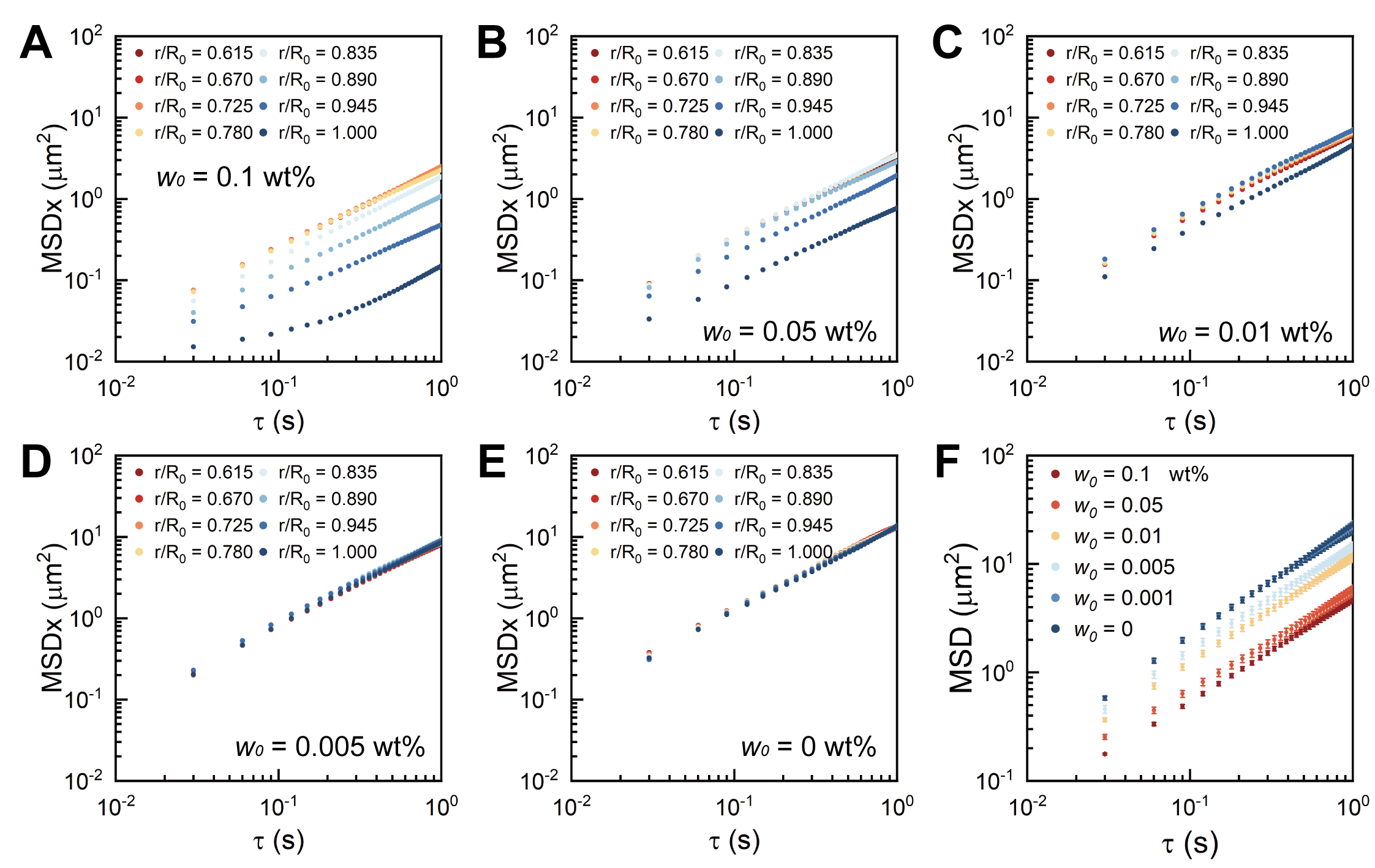}
\caption{Measurement of local viscosity and concentration during evaporation.
(\textit{A})-(\textit{E}) Local MSD$_\text{x}$ for tracer particle trajectories at different normalized radius regions, $r/R_0$, versus lag time, $\tau$, inside CMC-Na aqueous droplet with different initial concentrations.
(\textit{F}) MSD for tracer particle trajectories versus lag time, $\tau$, of CMC-Na aqueous solution with different concentrations.} 
\label{Micro-rheology}
\end{figure*}

\begin{figure*}[h!]
\centering
\includegraphics[width=0.5\textwidth]{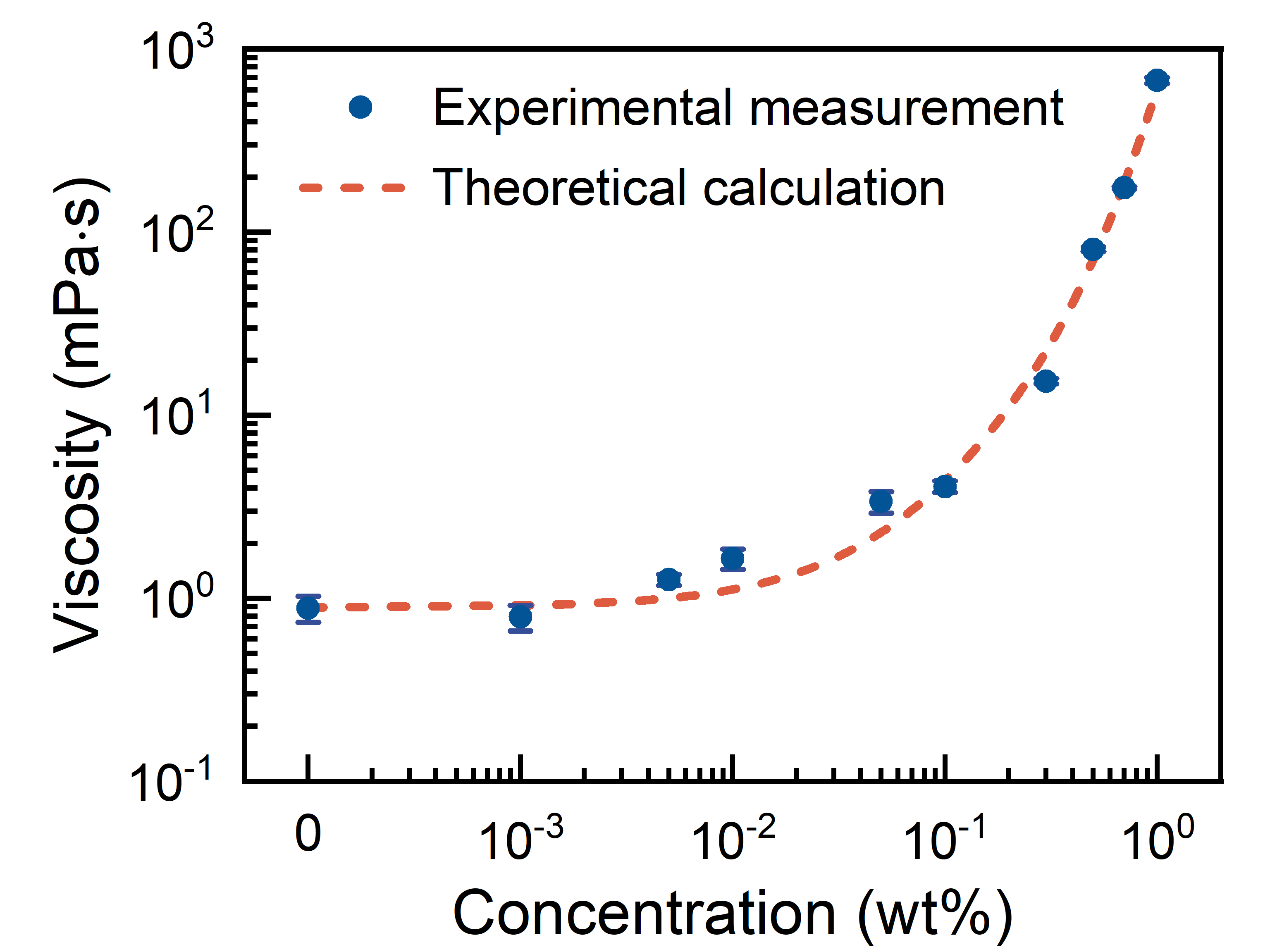 }
\caption{Viscosity of CMC-Na aqueous with initial concentrations ranging from 0.1 wt$\%$ to 1.0 wt$\%$. The blue dots indicate experimental measurement results measured by micro-rheology and viscometer. The orange dashed line indicates theoretical calculation results from previous reported work.}
\label{Viscosity to Concentration}
\end{figure*}

\begin{figure*}[h!]
\centering
\includegraphics[width=0.95\textwidth]{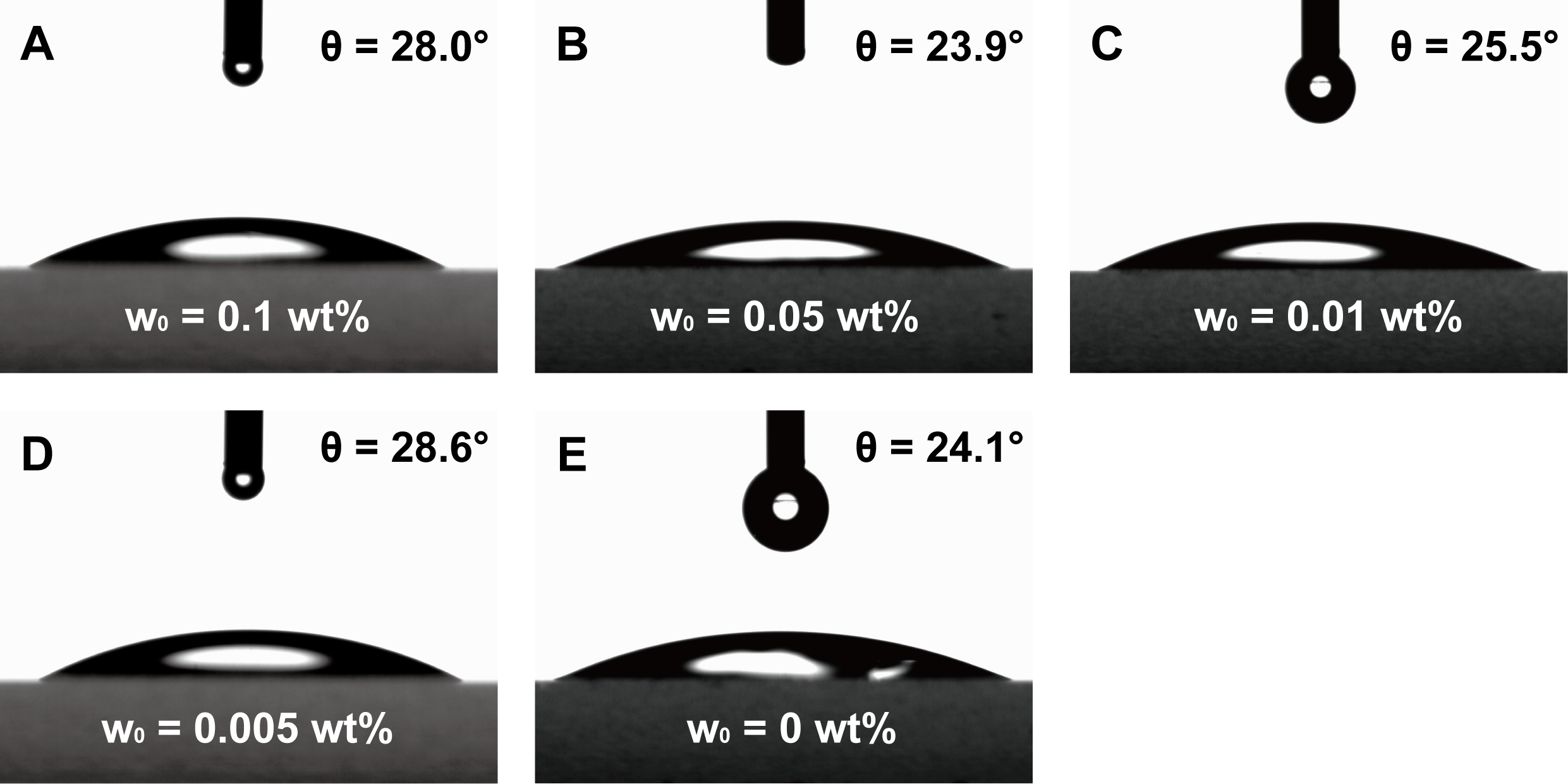}
\caption{The contact angle, $\theta$, of CMC-Na aqueous droplets with the concentrations, $w_0$, ranging from 0 wt$\%$ to 0.1 wt$\%$ on the glass slide.
For every measurement, a 2.5 $\upmu$L droplet was deposited on the glass slide. The contact angle was recorded by using a Dataohysics OCA20 contact angle measuring system at ambient temperature.
The measurement results indicate that there is no significant change in the contact angle of the solution after adding CMC-Na with concentrations ranging from 0 wt$\%$ to 0.1 wt$\%$.
Therefore, in our theoretical calculations, we maintain a consistent set of contact angles, i.e. $\theta=0.45$, for droplets of CMC-Na aqueous solution across all concentrations}
\label{Contact angle}
\end{figure*}

\twocolumngrid
\clearpage
\bibliography{apssamp}

\end{document}